\documentclass[12pt]{JHEP3}

\usepackage{cite}
\usepackage{dsfont}
\usepackage{amsfonts}
\usepackage{amssymb,latexsym}
\usepackage{amsmath,amsbsy}
\usepackage{epsfig,bm}
\usepackage{graphicx,comment}
\usepackage[active]{srcltx}

\def\cE{{\cal E}}   
   
  \def\cO{{\cal O}}

\def\tr{\mathop{\rm tr}}

\newcommand{\tmfloatcontents}{}
\newlength{\tmfloatwidth}
\newcommand{\tmfloat}[5]{
  \renewcommand{\tmfloatcontents}{#4}
  \setlength{\tmfloatwidth}{\widthof{\tmfloatcontents}+1in}
  \ifthenelse{\equal{#2}{small}}
    {\ifthenelse{\lengthtest{\tmfloatwidth > \linewidth}}
      {\setlength{\tmfloatwidth}{\linewidth}}{}}
    {\setlength{\tmfloatwidth}{\linewidth}}  \begin{minipage}[#1]{\tmfloatwidth}
    \begin{center}
      \tmfloatcontents
      \captionof{#3}{#5}
    \end{center}
  \end{minipage}}

\newcommand{\be}{\begin{equation}}
\newcommand{\ee}{\end{equation}}
\newcommand{\bea}{\begin{eqnarray}}
\newcommand{\eea}{\end{eqnarray}}
\newcommand{\bem}{\begin{multline}}
\newcommand{\eem}{\end{multline}}
\newcommand{\beg}{\begin{gather}}
\newcommand{\eeg}{\end{gather}}

\def\eq#1{{Eq.~(\ref{#1})}}
\def\fig#1{{Fig.~\ref{#1}}}
\newcommand{\ben}{\begin{eqnarray*}}
\newcommand{\een}{\end{eqnarray*}}
\newcommand{\un}[1]{\underline{#1}}

\def\peq#1{{(\ref{#1})}}

\graphicspath{{Figs/}}

\title{Long-Range Rapidity Correlations in Heavy Ion Collisions at
  Strong Coupling from AdS/CFT}

\author{Hovhannes R. Grigoryan, Yuri V. Kovchegov \\
\vspace{0.1in}

Department of Physics, The Ohio State University, Columbus,
OH 43210, USA \\~~\\

E-mail addresses: \email{grigoryan@physics.osu.edu},
\email{kovchegov.1@asc.ohio-state.edu}

\vspace{0.1in}
}

\abstract{We use AdS/CFT correspondence to study two-particle
  correlations in heavy ion collisions at strong coupling.  Modeling
  the colliding heavy ions by shock waves on the gravity side, we
  observe that at early times after the collision there are long-range
  rapidity correlations present in the two-point functions for the
  glueball and the energy-momentum tensor operators. We estimate
  rapidity correlations at later times by assuming that the evolution
  of the system is governed by ideal Bjorken hydrodynamics, and find
  that glueball correlations in this state are suppressed at large
  rapidity intervals, suggesting that late-time medium dynamics can
  not ``wash out'' the long-range rapidity correlations that were
  formed at early times.  These results may provide an
  insight on the nature of the ``ridge'' correlations observed in
  heavy ion collision experiments at RHIC and LHC, and in
  proton-proton collisions at LHC.}


\preprint{\today}

\begin{document}

\section{Introduction}

In recent years it has been suggested that the medium of quarks and
gluons produced in heavy ion collisions at RHIC goes through a
strongly-coupled phase at least during some period of its evolution
\cite{Teaney:2003kp,Shuryak:2006se,Huovinen:2001cy,Teaney:2001av}. The
Anti-de Sitter space/Conformal Field Theory (AdS/CFT) correspondence
\cite{Maldacena:1997re,Gubser:1998bc,Witten:1998qj} is often used to
study the dynamics of this strongly-coupled medium
\cite{Janik:2005zt,Janik:2006ft,Heller:2007qt,Benincasa:2007tp,Kovchegov:2007pq,Kajantie:2008rx,Grumiller:2008va,Gubser:2008pc,Albacete:2008vs,Albacete:2009ji,Lin:2009pn,Gubser:2009sx,AlvarezGaume:2008fx,Nastase:2008hw,Kovchegov:2009du,Taliotis:2010pi,Lin:2010cb,Chesler:2009cy,Gubser:2010ze,Chesler:2010bi,Beuf:2009cx,Beuf:2008ep}:
while it is valid only for ${\cal N} =4$ super-Yang-Mills (SYM)
theory, there is a possibility that the qualitative (and some of the
quantitative) results obtained from AdS/CFT correspondence may be
applied to the real-world case of QCD.

The main thrust of the efforts to study the dynamics of the medium
produced in heavy ion collisions using AdS/CFT correspondence has been
directed toward understanding how (and when) the medium isotropizes
and thermalizes
\cite{Janik:2005zt,Janik:2006ft,Heller:2007qt,Benincasa:2007tp,Kovchegov:2007pq,Kajantie:2008rx,Grumiller:2008va,Gubser:2008pc,Albacete:2008vs,Albacete:2009ji,Lin:2009pn,Gubser:2009sx,AlvarezGaume:2008fx,Nastase:2008hw,Kovchegov:2009du,Taliotis:2010pi,Lin:2010cb,Chesler:2009cy,Chesler:2010bi,Beuf:2009cx}.
The existing approaches can be divided into two categories: while some
studies concentrated on the dynamics of the produced medium in the
forward light-cone without analyzing the production mechanism for the
medium
\cite{Janik:2005zt,Kovchegov:2007pq,Janik:2006ft,Heller:2007qt,Benincasa:2007tp,Chesler:2009cy,Beuf:2009cx},
a large amount of work has been concentrated on studying the
collisions by modeling the heavy ions with shock waves in AdS$_5$ and
attempting to solve Einstein equations in the bulk for a collision of
two AdS$_5$ shock waves
\cite{Kajantie:2008rx,Grumiller:2008va,Gubser:2008pc,Albacete:2008vs,Albacete:2009ji,Lin:2009pn,Gubser:2009sx,AlvarezGaume:2008fx,Nastase:2008hw,Kovchegov:2009du,Taliotis:2010pi,Lin:2010cb,Chesler:2010bi}.
Many of the existing calculations strive to obtain the expectation
value of the energy-momentum tensor $\langle T_{\mu\nu} \rangle$ of
the produced medium in the boundary gauge theory
\cite{Janik:2005zt,Janik:2006ft,Heller:2007qt,Benincasa:2007tp,Kovchegov:2007pq,Kajantie:2008rx,Grumiller:2008va,Albacete:2008vs,Albacete:2009ji,Taliotis:2010pi,Lin:2010cb,Chesler:2009cy,Chesler:2010bi,Beuf:2009cx},
since this is the quantity most relevant for addressing the question
of the isotropization of the medium. Other works address the general
question of thermalization by noticing that it corresponds to creation
of a black hole in the AdS bulk, and by constructing a physical proof
of the black hole formation with the help of a trapped surface
analysis
\cite{Gubser:2008pc,Lin:2009pn,Gubser:2009sx,AlvarezGaume:2008fx,Nastase:2008hw,Kovchegov:2009du}.

In this work we concentrate on a different observable characterizing
heavy ion collisions: we study correlation functions in the produced
expanding strongly-coupled medium. Correlation functions have become a
powerful tool for the analysis of data coming out of heavy ion
collisions, allowing for a quantitative measure of a wide range of
phenomena, from Hanbury-Brown--Twiss (HBT) interferometry
\cite{Adler:2001zd}, to jet quenching \cite{Adler:2002tq} and Color
Glass Condensate (CGC) \cite{Braidot:2010ig}. In recent years a new
puzzling phenomenon was discovered in the two-particle correlation
functions measured in $Au+Au$ collisions at Relativistic Heavy Ion
Collider (RHIC) \cite{Adams:2005ph,Adare:2008cqb,Alver:2009id}: the
experiments see correlations with a rather small azimuthal angle
spread, but with a rather broad (up to several units) distribution in
rapidity. This type of correlation is referred to as ``the ridge''.
More recently the ridge correlations have been seen in
high-multiplicity proton-proton collisions at the Large Hadron
Collider (LHC) \cite{Khachatryan:2010gv}, as well as in the
preliminary data on $Pb+Pb$ collisions at LHC.

\begin{figure}[th]
\begin{center}
\epsfxsize=8cm
\leavevmode
\hbox{\epsffile{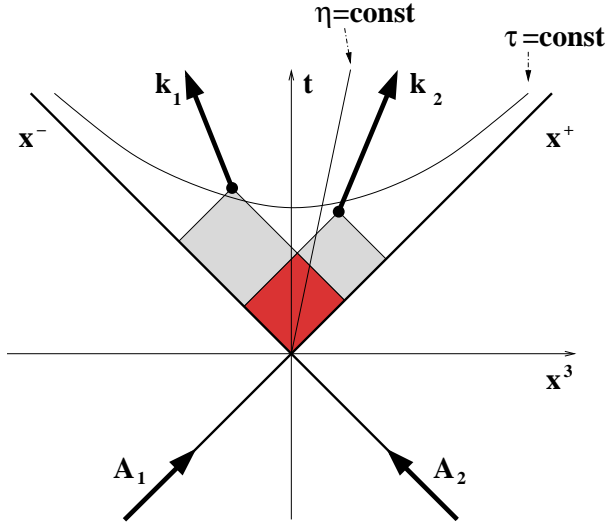}}
\end{center}
\caption{Space-time picture of a heavy ion collision demonstrating
  how long-range rapidity correlations can be formed only in the
  initial stages of the collision, as originally pointed out in
  \cite{Gavin:2008ev,Dumitru:2008wn}. Gray shaded regions denote
  causal pasts of the two produced particles with four-momenta $k_1$
  and $k_2$, with their overlap region highlighted in red. We have
  drawn the lines of constant proper time $\tau$ and constant
  space-time rapidity $\eta$ to guide the eye and to underscore that
  late-time emission events for the two particles are likely to be
  causally disconnected. }
\label{spacetime}
\end{figure}

Several theoretical explanations have been put forward to account for
the ridge correlations. They can be sub-divided into two classes:
perturbative and non-perturbative. Perturbative explanations, put
forward in the CGC framework in
\cite{Dumitru:2008wn,Gavin:2008ev,Dusling:2009ni,Dumitru:2010iy,Dumitru:2010mv,Kovner:2010xk},
are based on the long-range rapidity correlations present in the
initial state of a heavy ion collision due to CGC classical gluon
fields
\cite{McLerran:1993ni,Kovner:1995ja,Kovchegov:1997ke,Kovchegov:1999ep}
(see \cite{Jalilian-Marian:2005jf,Weigert:2005us,Iancu:2003xm} for
reviews of CGC physics). In \cite{Gavin:2008ev,Dumitru:2008wn} the
authors invoke causality to argue that long-range rapidity correlation
can only arise in the early times after the collision, since at later
times the regions at different rapidities become causally
disconnected. This is illustrated in \fig{spacetime}, where one can
see that the gray-shaded causal pasts of two particles produced in the
collision (labeled by arrows with momenta $k_1$ and $k_2$) overlap
only at very early time (the red-shaded region).  The authors of
\cite{Gavin:2008ev} then suggest that the late-time radial flow due to
hydrodynamic evolution would lead to azimuthal correlations
characteristic of the ``ridge''.  Alternatively, the authors of
\cite{Dumitru:2008wn,Dusling:2009ni,Dumitru:2010mv,Dumitru:2010iy}
have identified a class of Feynman diagrams which generate azimuthal
correlations in nucleus--nucleus collisions.

The CGC correlations found in
\cite{Dumitru:2008wn,Dusling:2009ni,Dumitru:2010mv,Dumitru:2010iy} are
based on purely perturbative small-coupling physics: however, it
remains to be shown whether such perturbative dynamics contains large
enough azimuthal correlations to account for all of the observed
``ridge'' phenomenon. In the scenario of
\cite{Gavin:2008ev,Dumitru:2008wn} CGC dynamics provides rapidity
correlations, while azimuthal correlations are generated by
hydrodynamic evolution. As we have already mentioned, it is possible
that the medium created at RHIC is strongly-coupled
\cite{Teaney:2003kp,Shuryak:2006se,Huovinen:2001cy,Teaney:2001av}: if
so, hydrodynamic evolution would then be a non-perturbative effect,
making the scenario proposed in \cite{Gavin:2008ev,Dumitru:2008wn}
implicitly non-perturbative. Purely non-perturbative explanations of
the ``ridge'' include parton cascade models \cite{Werner:2010ss},
hadronic string models \cite{Konchakovski:2008cf}, and event-by-event
hydrodynamic simulations \cite{Takahashi:2009na}. The causality
argument of \cite{Gavin:2008ev,Dumitru:2008wn} is valid in the
non-perturbative case as well: one needs correlations in the initial
state, either due to soft pomeron/hadronic strings interactions
\cite{Werner:2010ss,Konchakovski:2008cf}, or due to initial-state
fluctuations \cite{Takahashi:2009na}, in order to obtain long-range
rapidity correlations. In this work we will use AdS/CFT to address the
theoretical question whether long-range rapidity correlations are
present in the non-perturbative picture of heavy ion collisions.  At
the same time we recognize that a complete understanding of whether
the ``ridge'' correlations observed at RHIC and LHC are perturbative
(CGC) or non-perturbative in nature is still an open problem left for
future studies.

The goal of the present work is to study long-range rapidity
correlations in heavy ion collisions in the strongly-coupled AdS/CFT
framework. In order to test for the long range rapidity correlations
observed in heavy ion collisions, we would like to study the two-point
function $\langle\tr F_{\mu\nu}^2(x) \, \tr
F_{\rho\sigma}^2(y)\rangle$ of glueball operators $\tr
\left(F_{\mu\nu}^2 \right)$ right after the collision but before the
thermalization. According to causality arguments of
\cite{Gavin:2008ev,Dumitru:2008wn}, one expects that the long range
correlations in rapidity should occur at such early times. The choice
of observable is mainly governed by calculational simplicity. The
metric for the early times after the collision of two shock waves in
AdS$_5$ was obtained in
\cite{Grumiller:2008va,Albacete:2008vs,Albacete:2009ji}: after
formulating the problem in Sec. \ref{general} and presenting our
general expectation for the answer in Sec. \ref{simple}, we use this
metric to calculate the correlation function of two glueball operators
in Sec. \ref{Correlators}. (Since the glueball operator corresponds to
the massless scalar field in the bulk, we compute the two-point
function of the scalar field in the background of the colliding shock
waves metric.)  Our main result is that we do find long-range rapidity
correlations in the strongly-coupled initial state, albeit with a
rather peculiar rapidity dependence: the two-glueball correlation
function scales as
\begin{align}
  \label{eq:ampl}
  C (k_1, k_2) \, \sim \, \cosh \left( 4 \, \Delta y \right)
\end{align}
with the (large) rapidity interval $\Delta y$ between them. We also
show in Sec. \ref{Correlators} that the correlator of two
energy-momentum tensors $\langle T_{2}^1 (x) \, T_{2}^1 (y) \rangle$
(with $1,2$ transverse directions) exhibits the same long-range
rapidity correlations. This should be contrasted with the CGC result,
in which the correlations are at most flat in rapidity
\cite{Dumitru:2008wn,Dumitru:2010iy,Dumitru:2010mv,Gavin:2008ev}.
Indeed the growth of correlations with rapidity interval in
\eq{eq:ampl} also contradicts experimental data
\cite{Adams:2005ph,Adare:2008cqb,Alver:2009id,Khachatryan:2010gv}.
Although we should not {\it a priori} expect an agreement between
AdS/CFT calculations and experimental QCD data, we argue in Sec.
\ref{sum} that inclusion of higher-order corrections in the AdS
calculation along the lines of \cite{Albacete:2009ji} should help 
to flatten out such growth, though it is a very difficult problem to
demonstrate this explicitly.

Using the causality argument of \cite{Gavin:2008ev,Dumitru:2008wn}
illustrated in \fig{spacetime} we also expect that after
thermalization the rapidity correlations should only be short-ranged.
As a result, due to causality, the initial long ranged correlations
can not be ``washed away'' and will be observed at later times.
This explanation is analogous to the resolution of the `horizon
problem' in the cosmic microwave background radiation (CMB), where the
observed near-homogeneity of the CMB suggests that the universe was
extremely homogeneous at the time of the last scattering even over
distance scales that could not have been in causal contact in the
past.  This problem was solved by assuming that the universe, when it
was still young and extremely homogeneous, went through a very rapid
period of expansion (inflation).  As a consequence of inflation,
different regions of the universe became causally disconnected, while
preserving the initial homogeneity.
The idea that we pursue here for the heavy ion collisions seems to be
of similar nature. To verify the statement that late-time dynamics can
not generate (or otherwise affect) long-range rapidity correlations we
study glueball correlation again in Sec. \ref{late-times} now using
the metric found by Janik and Peschanski \cite{Janik:2005zt}, which is
dual to Bjorken hydrodynamics \cite{Bjorken:1982qr}. (This is done in
the absence an analytic solution of the problem of colliding shock
waves: despite some recent progress
\cite{Grumiller:2008va,Albacete:2008vs,Albacete:2009ji} the late-time
metric is unknown at present.) Performing a perturbative estimate, we
find that, indeed, only short-range rapidity correlations result from
the gauge theory dynamics dual to the Janik and Peschanski metric.

We summarize our results in Sec. \ref{sum}.


\section{Generalities and Problem Setup}
\label{general}

\subsection{AdS/CFT Tools}

We start with a metric for a single shock wave moving along a light
cone in the $x^+$ direction \cite{Janik:2005zt} in Fefferman--Graham
coordinates \cite{F-G}:
\begin{equation}\label{1nuc}
  ds^2 \, = \, \frac{L^2}{z^2} \, \left\{ -2 \, dx^+ \, dx^- + t_1 (x^-) \, z^4 \, d
    x^{- \, 2} + d x_\perp^2 + d z^2 \right\}
\end{equation}
where
\begin{align}\label{t1}
  t_1 (x^-) \, \equiv \, \frac{2 \, \pi^2}{N_c^2} \, \langle T_{1 \,
    --} (x^-) \rangle.
\end{align}
Here $x^\pm = \frac{x^0 \pm x^3}{\sqrt{2}}$, ${\un x} = (x^1, x^2)$,
$d x_\perp^2 = (d x^1)^2 + (d x^2)^2$, $z$ is the coordinate
describing the 5th dimension such that the ultraviolet (UV) boundary
of the AdS space is at $z=0$, and $L$ is the radius of the AdS space.
According to holographic renormalization \cite{deHaro:2000xn},
$\langle T_{--} (x^-) \rangle$ is the expectation value of the
energy-momentum tensor for a single ultrarelativistic nucleus moving
along the light-cone in the $x^+$-direction in the gauge theory. We
assume that the nucleus is made out of nucleons consisting of $N_c^2$
``valence gluons'' each, such that $\langle T_{--} (x^-) \rangle
\propto N_c^2$, and the metric \peq{1nuc} has no $N_c^2$-suppressed
terms in it. The metric in \eq{1nuc} is a solution of Einstein
equations in AdS$_5$:
\begin{align}
  \label{ein}
  R_{\mu\nu} + \frac{4}{L^2} \, g_{\mu\nu} = 0.
\end{align}

Imagine a collision of the shock wave \peq{1nuc} with another similar
shock wave moving in the light cone $x^-$ direction described by the
metric
\begin{align}\label{2nuc}
  ds^2 \, = \, \frac{L^2}{z^2} \, \left\{ -2 \, dx^+ \, dx^- + t_2
    (x^+) \, z^4 \, d x^{+ \, 2} + d x_\perp^2 + d z^2 \right\}
\end{align}
with
\begin{align}\label{t2}
  t_2 (x^+) \, \equiv \, \frac{2 \, \pi^2}{N_c^2} \, \langle T_{2 \,
    ++} (x^+) \rangle.
\end{align}
Here we will consider the high-energy approximation, in which the
shock waves' profiles are given by delta-functions,
\begin{align}\label{deltas}
  t_1 (x^-) = \mu_1 \, \delta (x^-), \ \ \ t_2 (x^+) = \mu_2 \, \delta
  (x^+).
\end{align}
The two scales $\mu_1$ and $\mu_2$ can be expressed in terms of the
physical parameters in the problem since we picture the shock waves as
dual to the ultrarelativistic heavy ions in the boundary gauge theory
\cite{Albacete:2008vs,Albacete:2008ze} :
\begin{align}\label{mus}
  \mu_{1} \sim p_{1}^+ \, \Lambda_1^2 \, A_1^{1/3}, \ \ \ \mu_{2} \sim
  p_{2}^- \, \Lambda_2^2 \, A_2^{1/3}.
\end{align}
Here $p_1^+$, $p_2^-$ are the large light-cone momenta per nucleon,
$A_1$ and $A_2$ are atomic numbers, and $\Lambda_1$ and $\Lambda_2$
are the typical transverse momentum scales in the two nuclei
\cite{Albacete:2008vs}. Note that $\mu_1$ and $\mu_2$ are independent
of $N_c$.

The exact analytical solution of Einstein equations \peq{ein} starting
with the superposition of the metrics \peq{1nuc} and \peq{2nuc} before
the collision, and generating the resulting non-trivial metric after
the collisions, is not known. Instead one constructs perturbative
expansion of the solution of Einstein equations in powers of $t_1$ and
$t_2$, or, equivalently, $\mu_1$ and $\mu_2$
\cite{Grumiller:2008va,Albacete:2008vs,Albacete:2009ji,Taliotis:2010pi,Lin:2010cb}.
At present the metric is known up to the fourth order in $\mu$'s
\cite{Grumiller:2008va,Albacete:2008vs,Albacete:2009ji}, and also a
resummation to all-orders in $\mu_2$ ($\mu_1$) while keeping $\mu_1$
($\mu_2$) at the lowest order has been performed in
\cite{Albacete:2009ji}. The validity of the perturbatively obtained
metric is limited to early proper times $\tau = \sqrt{2 \, x^+ \,
  x^-}$, see e.g. \cite{Albacete:2009ji} (though indeed the
fully-resummed series in powers of $\mu_1$, $\mu_2$ would be valid
everywhere). Since here we are interested in the early-time
correlations (and due to complexity of the $\mu_2$-resummed metric
obtained in \cite{Albacete:2009ji}), we limit ourselves to the $O
(\mu_1 \, \mu_2)$ metric obtained in
\cite{Albacete:2008vs,Grumiller:2008va} in the Fefferman--Graham
coordinates:
\begin{align}\label{2nuc_gen}
  ds^2 \, = \, \frac{L^2}{z^2} \, \bigg\{ -\left[ 2 + G (x^+, x^-, z)
  \right] \, dx^+ \, dx^- + \left[ t_1 (x^-) \, z^4 + F (x^+, x^-, z)
  \right] \, d x^{- \, 2} \notag \\ + \left[ t_2 (x^+) \, z^4 +
    {\tilde F} (x^+, x^-, z) \right] \, d x^{+ \, 2} + \left[ 1 + H
    (x^+, x^-, z) \right] \, d x_\perp^2 + d z^2 \bigg\}.
\end{align}
The components of the metric at the order-$\mu_1 \, \mu_2$ are
\begin{align}\label{LO}
  F (x^+, x^-, z) \, & = \, - \lambda_1 (x^+, x^-) \, z^4 -
  \frac{1}{6} \, \partial_-^2 h_0 (x^+, x^-) \, z^6 - \frac{1}{16} \,
  \partial_-^2
  h_1 (x^+, x^-) \, z^8 \notag \\
  {\tilde F} (x^+, x^-, z) \, & = \, - \lambda_2 (x^+, x^-) \, z^4 -
  \frac{1}{6} \, \partial_+^2 h_0 (x^+, x^-) \, z^6 - \frac{1}{16} \,
  \partial_+^2
  h_1 (x^+, x^-) \, z^8 \notag \\
  G (x^+, x^-, z) \, & = \, - 2 \, h_0 (x^+, x^-) \, z^4 - 2 \, h_1
  (x^+, x^-) \, z^6 + \frac{2}{3} \, t_1 (x^-) \, t_2 (x^+) \, z^8 \notag \\
  H (x^+, x^-, z) \, & = \, h_0 (x^+, x^-) \, z^4 + h_1 (x^+, x^-) \,
  z^6,
\end{align}
where we defined \cite{Albacete:2008vs}
\begin{align}\label{LOstuff}
  h_0 (x^+, x^-) \, & = \, \frac{8}{\partial_+^2 \, \partial_-^2} \,
  t_1 (x^-) \, t_2 (x^+), \ \ \ h_1 (x^+, x^-) \, = \, \frac{4}{3 \,
    \partial_+ \, \partial_-} \, t_1 (x^-) \, t_2 (x^+) \notag \\
  \lambda_1 (x^+, x^-) \, & = \, \frac{\partial_{-}}{\partial_{+}} \,
  h_0 (x^+, x^-), \ \ \ \lambda_2 (x^+, x^-) \, = \,
  \frac{\partial_{+}}{\partial_{-}} \, h_0 (x^+, x^-)
\end{align}
along with the definition of the causal integrations
\begin{align}\label{ints}
  \frac{1}{\partial_{+}} [\ldots](x^+) \, \equiv \,
  \int\limits_{-\infty}^{x^+} \, d x'^+ \, [\ldots](x'^+), \ \ \
  \frac{1}{\partial_{-}} [\ldots](x^-) \, \equiv \,
  \int\limits_{-\infty}^{x^-} \, d x'^- \, [\ldots](x'^-).
\end{align}

Below we will calculate correlation functions of the glueball
operators
\begin{align}\label{Jdef}
  J(x) \, \equiv \, \frac{1}{2} \, \tr [F_{\mu\nu} \, F^{\mu\nu}]
\end{align}
in the boundary gauge theory.\footnote{When defining the glueball
  operator we assume that in the boundary theory the gluon field
  $A_\mu^a$ is defined without absorbing the gauge coupling $g_{YM}$
  in it, such that the field strength tensor $F_{\mu\nu}^a$ contains
  the coupling $g_{YM}$.} According to the standard AdS/CFT
prescription,\footnote{Since $\Delta =4$, with $\Delta$ the conformal
  dimension of $J(x)$ , the mass of the dual scalar field, $m^2 =
  \Delta \, (\Delta-4)$, is zero.} the glueball operator is dual to
the massless scalar (dilaton) field $\phi$ in the AdS$_5$ bulk
\cite{Klebanov:2000me} with the action
\begin{align}
  S^\phi \, = \, - \frac{N_c^2}{16 \, \pi^2 \, L^3} \, \int d^4 x \, d
  z \, \sqrt{-g} \, g^{MN} \, \partial_M \phi (x,z) \, \partial_N \phi
  (x,z),
\end{align}
where $M,N = (\mu,z)$, $\mu = (0,1,2,3)$ and $x^{\mu}$ correspond to
4D field theory coordinates, while $z$ is the coordinate along the
extra fifth (holographic) dimension. (As usual $g = \det {g_{MN}}$.)

The equation of motion (EOM) for the scalar field is
\begin{align}\label{eom}
  \frac{1}{\sqrt{-g}} \, \partial_M \left[ \sqrt{-g}~g^{MN}\partial_N
    \phi (x,z)\right] \, = \, 0.
\end{align}
Using \eq{eom}, the dilaton action evaluated on the classical solution
can be cast in the following form convenient for the calculation of
correlation functions:
\begin{align}
  \label{dil_action}
  S^\phi_{cl} \, = \, \frac{N_c^2}{16 \, \pi^2 \, L^3} \, \int d^4 x
  \, \left[ \sqrt{-g} \, g^{zz} \, \phi(x,z) \, \partial_z \phi(x,z)
  \right] \Bigg|_{z=0} \, = \, \frac{N_c^2}{16 \, \pi^2} \, \int d^4 x
  \, \phi_B (x) \, \left[ \frac{1}{z^3} \, \partial_z \phi(x,z)
  \right] \Bigg|_{z=0}.
\end{align}
In arriving at the expression on the right of \eq{dil_action} we have
used the metric in Eqs. \peq{2nuc_gen}, \peq{LO}, and \peq{LOstuff},
along with the standard assumption that the fields $\phi$ have the
following boundary condition (BC) at the UV boundary, $\phi(x,z\to 0)
= \phi_B(x)$, which allowed us to approximate near $z=0$
\begin{align}
  \label{g_eq}
  g \, = \, - \frac{L^{10}}{z^{10}} \, \left( 1 - \frac{1}{3} \, z^8
    \, t_1 (x^-) \, t_2 (x^+) \right) \, \approx \, -
  \frac{L^{10}}{z^{10}}.
\end{align}
In arriving at \eq{dil_action} we have also demanded that\footnote{As
  one can see later, our classical solutions satisfy this condition.}
\begin{align}
  \sqrt{-g} \, g^{zz} \, \phi(x,z) \, \partial_z \phi(x,z) \,
  \rightarrow \, 0 \ \ \ \text{as} \ \ \ z \rightarrow \infty.
\end{align}

Define the retarded Green function of the glueball operator \peq{Jdef}
(averaged in the heavy ion collision background),
\begin{align}
  \label{retG}
  G_R (x_1, z_2) \, = \, - i \, \theta (x_1^0 - x_2^0) \, \langle
  [J(x_1), J(x_2)] \rangle.
\end{align}
According to the AdS/CFT correspondence the contribution to the
retarded Green function coming from the medium produced in the
collision is given by \cite{Son:2002sd}\footnote{As was shown in
  \cite{Herzog:2002pc,Skenderis:2008dg} the right-hand side of
  \eq{Sdiff} contains contributions of both the retarded and advanced
  Green functions $G_R$ and $G_A$. In the lowest-order calculation we
  are going to perform here the Green functions are real, and, since
  $\text{Re} \, G_R = \text{Re} \, G_F = \text{Re} \, G_A$ (with $G_F$
  the Feynman Green function defined below in \eq{GF}), we do not need
  to address the question of disentangling the contributions of
  different wave functions to \eq{Sdiff} and will adopt the convention
  of \cite{Mueller:2008bt,Avsar:2009xf} by calling the object in
  \eq{Sdiff} a retarded Green function.}
\begin{align}
  \label{Sdiff}
  G_R (x_1, x_2) \, = \, \frac{\delta^2 [S^\phi_{cl} - S_0]}{\delta
    \phi_B (x_1) \, \delta \phi_B (x_2)},
\end{align}
where we subtract the action $S_0$ of the scalar field in the empty
AdS$_5$ space to remove the contribution of the retarded Green
function in the vacuum. The latter has nothing to do with the
properties of the medium produced in the collision and has to be
discarded.

Later we will be interested in the Fourier transform of the retarded
Green function
\begin{align}
  \label{Gr_mom}
  G_R (k_1, k_2) \, = \, \int d^4 x_1 \, d^4 x_2 \, e^{-i \, k_1 \cdot
    x_1 -i \, k_2 \cdot x_2} \, G_R (x_1, x_2).
\end{align}
(We are working in the $(-,+,+,+)$ metric in the boundary four
dimensions.)


\subsection{Kinematics}
\label{kine}

We have defined above $k^{\pm} = (k^0 \pm k^3)/\sqrt{2}$, ${\un k} =
(k^1, k^2)$, $k_\perp = |{\un k}|$ and $k^2 = k_{\bot}^2 - 2 \, k^+ \,
k^- \, = \, -m^2$. The particle rapidity, defined as, $y =
\frac{1}{2}\, \ln \frac{k^+}{k^-}$, is a useful variable, since the
rapidity difference between any pair of particles remains unchanged if
we go from the center of mass frame to any other frame by performing a
boost along the longitudinal direction, $x^3$.  On the other hand,
when $k^0 \gg m $, $y \approx y_p = \ln \cot (\theta/2)$, where $y_P$
is pseudorapidity, and $\theta$ is the angle at which the particle
emerges in the center of mass frame.  Furthermore, defining $m_{\bot}
\equiv \sqrt{k_{\bot}^2+m^2}$, we can rewrite the light-cone
components of the momentum as: $k^+ = m_{\bot}e^y/\sqrt{2}$ and $k^- =
m_{\bot}e^{-y}/\sqrt{2}$.  In the case when $k_{\bot}^2 \gg m^2$ one
has $k^+k^- \approx k^2_{\bot}/2$.

Consider two identical on mass-shell particles with momenta $k_1 =
(k_1^+, k^-_1, {\un k}_{1})$ and $k_2 = (k_2^+,k^-_2, {\un k}_{2})$.
Assuming $k^2_1 = k^2_2 = -m^2$ and ${\un k}_{1} = {\un k}_{2} = {\un
  k}$, we obtain
\begin{align}
  q^2 \equiv (k_2-k_1)^2 = -2 \, m^2 -2 \, k_1 \cdot k_2 \, = \,
  4 \, m^2_{\bot} \, \sinh^2 \frac{\Delta y}{2} > 0 \ ,
\end{align}
where $\Delta y = y_2 - y_1$ with $y_1$ and $y_2$ the rapidities of
the two particles. In case when $k^2_{\bot} \gg m^2$ and $\Delta y \gg
1$, we have $q^2 \approx 2 \, k_{\bot}^2 \cosh \Delta y \approx
k_{\bot}^2 e^{\Delta y}$. It is worth noting that the momentum
difference is space-like, since $q^2 \equiv Q^2 > 0$.


\subsection{Defining the observable in the boundary gauge theory}

Let us now specify the observable we want to calculate in the boundary
gauge theory. Our primary goal is to study rapidity correlations using
AdS/CFT. Ideally one would like to find correlations between produced
particles. However, ${\mathcal N} =4$ SYM theory has no bound states,
and, at strong coupling, it does not make sense to talk about
individual supersymmetric particles. Therefore we will study
correlators of operators, starting with the glueball operator defined
in \eq{Jdef}.  One can think of the glueballs as external probes to
${\mathcal N} =4$ SYM theory (in the sense of being particles from
some other theory in four dimensions), which couple to the gluons in
${\mathcal N} =4$ SYM, and therefore can be produced in the collision.
Later on we will also consider correlators of the energy-momentum
tensor $T_{\mu\nu}$, which should be also thought of as an operator
coupling to a particle (in four dimensions) external to the ${\mathcal
  N} =4$ SYM theory.

We start with the glueball production. To study two-particle
correlations we need to find the two-particle multiplicity
distribution
\begin{align}
  \label{N2}
  \frac{d^6 N}{d^2 k_1 \, d y_1 \, d^2 k_2 \, d y_2}
\end{align}
where $k_{1}^{\perp}$, $y_1$ and $k_{2}^{\perp}$, $y_2$ are the
transverse momenta of the produced particles (glueballs) and their
rapidities, and $d^2 k \equiv d k^1 \, d k^2$. As usual we can
decompose the two-particle multiplicity distribution into the
uncorrelated and correlated pieces
\begin{align}
  \label{2terms}
  \frac{d^6 N}{d^2 k_1 \, d y_1 \, d^2 k_2 \, d y_2} \, = \, \frac{d^3
    N}{d^2 k_1 \, d y_1} \, \frac{d^3 N}{d^2 k_2 \, d y_2} + \frac{d^6
    N_{corr}}{d^2 k_1 \, d y_1 \, d^2 k_2 \, d y_2}.
\end{align}
We are interested in computing the second (correlated) term on the
right hand side of \peq{2terms}. We begin by writing it as
\begin{align}
  \label{ampl^2}
  \frac{d^6 N_{corr}}{d^2 k_1 \, d y_1 \, d^2 k_2 \, d y_2} \, \propto
  \, \langle | M (k_1, k_2) |^2 \rangle
\end{align}
where $M (k_1, k_2)$ is the two-particle production amplitude. (Note
that since we are primarily interested in rapidity dependence of
correlators, we are not keeping track of prefactors and other
coefficients not containing two-particle correlations.)

For the correlated term in \eq{2terms} the amplitude of inclusive
two-glueball production in a heavy ion collision is
\begin{align}
  \label{ampl1}
  M (k_1, k_2) \, \propto \, \int d^4 x_1 \, d^4 x_2 \, e^{- i \, k_1
    \cdot x_1 - i \, k_2 \cdot x_2} \, \langle n | \, T \left\{ J
    (x_1) \, J (x_2) \right\} | A_1, A_2 \rangle,
\end{align}
which is a consequence of the LSZ reduction formula with $T$ denoting
time-ordering. Here $| n \rangle$ denotes an arbitrary state of the
gauge theory which describes other particles which may be produced in
a collision apart from the two glueballs.

The state $| A_1, A_2 \rangle$ can be thought of as the vacuum in the
presence of a source, with the source being the two nuclei with atomic
numbers $A_1$ and $A_2$.  Consider first the expectation value of the
energy-momentum operator $\langle T_{\mu\nu} \rangle$ in a nuclear
collision.  According to the standard prescription we can write it as
\begin{align}
  \label{Tmn}
  \langle T_{\mu\nu} (x) \rangle \, = \, \frac{\int {\cal D} A_\mu \,
    e^{i \, S [A]} \, W_+ [A] \, W_- [A] \, T_{\mu\nu} (x)}{\int {\cal
      D} A_\mu \, e^{i \, S [A]} \, W_+ [A] \, W_- [A]}
\end{align}
where $S [A]$ is the action of the gauge theory. For simplicity we
only explicitly show the integrals over gauge fields in \eq{Tmn},
implying the integrals over all other fields in the theory.  The
objects $W_+ [A]$ and $W_- [A]$ are some functionals of the fields in the theory describing the
two colliding nuclei. For instance, in the perturbative QCD approaches
such as CGC, these operators are Wilson lines along $x^-=0$ and $x^+
=0$ light cone directions
\cite{McLerran:1993ni,Kovner:1995ja,Kovchegov:1997ke,Kovchegov:1999ep}.
\footnote{Calculation of the expectation value of $T_{\mu\nu}$ in CGC
  is reduced to perturbative evaluation/resummation of \eq{Tmn} (see
  e.g.  \cite{Kovchegov:2005az} for an example of such calculation).}

Using operators and states in Heisenberg picture one can rewrite
\eq{Tmn} as
\begin{align}
  \label{eq:ave}
  \langle T_{\mu\nu} (x) \rangle \, = \, \langle A_1, A_2 | T_{\mu\nu}
  (x) |A_1, A_2 \rangle.
\end{align}
Comparing \eq{eq:ave} to \eq{Tmn} clarifies the meaning of the $| A_1,
A_2 \rangle$ state by demonstrating that the averaging in \eq{eq:ave}
is over a state of vacuum in the presence of nuclear sources (which of
course strongly disturb the vacuum).

Using \eq{ampl1} in \eq{ampl^2} we obtain
\begin{align}
  \label{corr2}
  \frac{d^6 N_{corr}}{d^2 k_1 \, d y_1 \, d^2 k_2 \, d y_2} \, \propto
  \, \int d^4 x_1 \, d^4 x_2 \, d^4 x'_1 \, d^4 x'_2 \, e^{- i \, k_1
    \cdot (x_1 - x'_1) - i \, k_2 \cdot (x_2 - x'_2)} \notag \\ \times
  \, \sum_n \, \langle A_1, A_2 | \, {\overline T} \left\{ J (x'_1) \,
    J (x'_2) \right\} | n \rangle \ \langle n | \, T \left\{ J (x_1)
    \, J (x_2) \right\} | A_1, A_2 \rangle
\end{align}
where ${\overline T}$ denotes the inverse time-ordering and we have
used the fact that $J (x)$ is a hermitean operator.  Summing over a
complete set of states $| n \rangle$ yields
\begin{align}
  \label{corr3}
  \frac{d^6 N_{corr}}{d^2 k_1 \, d y_1 \, d^2 k_2 \, d y_2} \, &
  \propto \, \int d^4 x_1 \, d^4 x_2 \, d^4 x'_1 \, d^4 x'_2 \, e^{- i
    \, k_1 \cdot (x_1 - x'_1) - i \, k_2 \cdot (x_2 - x'_2)} \notag \\
  & \times \, \langle A_1, A_2 | \, {\overline T} \left\{ J (x'_1) \,
    J (x'_2) \right\} \, T \left\{ J (x_1) \, J (x_2) \right\} | A_1,
  A_2 \rangle.
\end{align}

As one could have expected, in order to calculate two-particle
production, we need to calculate a 4-point function given in
\eq{corr3}. This is, in general, a difficult task: instead we will use
the following simplification. Begin by replacing the complete set of
states $| n \rangle$ by states ${\cal O}_n (x) \, |A_1, A_2 \rangle$
obtained by acting on our ``vacuum'' state $| A_1, A_2 \rangle$ by a
complete orthonormal set of gauge theory operators ${\cal O}_n (x)$,
such that
\begin{align}
  \label{unity}
  \mathds{1} \, = \, \sum_n \, | n \rangle \ \langle n | \, \, = \,
  \sum_n \, \int \, d^4 x \, {\cal O}_n (x) \, | A_1, A_2 \rangle \ \,
  \langle A_1, A_2 | \, {\cal O}_n^\dagger (x)
\end{align}
with the normalization condition
\begin{align}
  \label{norm}
  \langle A_1, A_2 | \, {\cal O}_m^\dagger (y) \, {\cal O}_n (x) \, |
  A_1, A_2 \rangle \, = \, \delta_{nm} \, \delta^{(4)} (x-y).
\end{align}
Using \eq{unity} in \eq{corr2} we write
\begin{align}
  \label{corr4}
  & \frac{d^6 N_{corr}}{d^2 k_1 \, d y_1 \, d^2 k_2 \, d y_2} \,
  \propto \, \int d^4 x_1 \, d^4 x_2 \, d^4 x'_1 \, d^4 x'_2 \, e^{- i
    \, k_1 \cdot (x_1 - x'_1) - i \, k_2 \cdot (x_2 - x'_2)} \, \sum_n
  \, \int \, d^4 x \notag \\ & \times \, \langle A_1, A_2 | \,
  {\overline T} \left\{ J (x'_1) \, J (x'_2) \, {\cal O}_n (x)
  \right\} \, | A_1, A_2 \rangle \ \, \langle A_1, A_2 | \, T \left\{
    {\cal O}_n^\dagger (x) \, J (x_1) \, J (x_2) \right\} | A_1, A_2
  \rangle.
\end{align}
To evaluate \eq{corr4} we have to insert all possible operators ${\cal
  O}_n (x)$ from the orthonormal set in it. Noting that $J (x)$ is a
gauge-invariant color-singlet operator, we conclude that only
color-singlet ${\cal O}_n (x)$ would contribute. Also, since the final
state in a scattering problem should be an observable, the operators
${\cal O}_n$ should be hermitean. The set of contributing ${\cal O}_n
(x)$'s should therefore include the identity operator, $J(x)$,
$T_{\mu\nu} (x)$, etc.

As we will see below, since we are using the metric \peq{2nuc_gen},
which is a perturbative solution of Einstein equations to order $\mu_1
\, \mu_2$, we can only calculate correlators to order $\mu_1 \, \mu_2$
as well. Moreover, correlators which are independent of $\mu_1$ and
$\mu_2$ are vacuum correlators that we are not interested in.
Correlators of order $\mu_1$ or $\mu_2$ correspond to performing deep
inelastic scattering (DIS) on a single shock wave similar to
\cite{Mueller:2008bt,Avsar:2009xf,Kovchegov:2010uk}, and are thus not
directly relevant to the problem of heavy ion collisions at hand. Thus
in this paper we are only interested in correlators exactly at the
order $\mu_1 \, \mu_2$ in the expansion in the two shock waves. Using
such power counting it is easy to see that inserting the identity
operator (normalized to one to satisfy \eq{norm}) into \eq{corr4} in
place of ${\cal O}_n$'s would give us a contribution of the order of
$\mu_1^2 \, \mu_2^2$, which is the lowest order contribution to double
glueball production. Inserting $J(x)$ or $T_{\mu\nu} (x)$ into
\eq{corr4} instead of ${\cal O}_n$'s would give zero. One can also see
that replacing ${\cal O}_n$'s by higher (even) powers of $J(x)$ or
$T_{\mu\nu} (x)$ (properly orthogonalized) in \eq{corr4} would
generate non-zero contributions, which are either higher order in
$\mu_1$ and $\mu_2$ or $N_c^2$-suppressed. We therefore insert the
identity operator into \eq{corr4}, which in the color space can be
written as ${\bf 1} = \delta^{ab}/N_c$ to satisfy normalization in
\eq{norm}, and write
\begin{align}
  \label{corr5}
  & \frac{d^6 N_{corr}}{d^2 k_1 \, d y_1 \, d^2 k_2 \, d y_2} \,
  \propto \, \int d^4 x_1 \, d^4 x_2 \, d^4 x'_1 \, d^4 x'_2 \, e^{- i
    \, k_1 \cdot (x_1 - x'_1) - i \, k_2 \cdot (x_2 - x'_2)} \notag \\
  & \times \, \frac{1}{N_c^2} \, \langle A_1, A_2 | \, {\overline T}
  \left\{ J (x'_1) \, J (x'_2) \right\} | A_1, A_2 \rangle \ \,
  \langle A_1, A_2 | \, T \left\{ J (x_1) \, J (x_2) \right\} | A_1,
  A_2 \rangle \, \left[1 + O (1/N_c^2) \right].
\end{align}
We have thus reduced the problem of two-glueball production to
calculation of two-point correlation functions! Note that the
prefactor of $1/N_c^2$ makes the $N_c$ counting right: since each
connected correlator is order-$N_c^2$, we see from \eq{corr5} that the
correlated two-particle multiplicity scales as $N_c^2$ as well, in
agreement with perturbative calculations
\cite{Dumitru:2008wn,Gavin:2008ev,Dusling:2009ni,Dumitru:2010iy,Dumitru:2010mv}.

Defining Feynman Green function
\begin{align}
  \label{GF}
  G_F (k_1, k_2) \, = \, \int d^4 x_1 \, d^4 x_2 \, e^{- i \, k_1
    \cdot x_1 - i \, k_2 \cdot x_2} \, \langle A_1, A_2 | \, T \left\{
    J (x_1) \, J (x_2) \right\} | A_1, A_2 \rangle
\end{align}
we can summarize \eq{corr5} as
\begin{align}
  \label{corr6}
  \frac{d^6 N_{corr}}{d^2 k_1 \, d y_1 \, d^2 k_2 \, d y_2} \, \propto
  \, \frac{1}{N_c^2} \, |G_F (k_1, k_2)|^2.
\end{align}

With the help of the retarded Green function
 \begin{align}
   \label{GR}
   G_R (k_1, k_2) \, = \, - i \, \int d^4 x_1 \, d^4 x_2 \, e^{- i \,
     k_1 \cdot x_1 - i \, k_2 \cdot x_2} \, \theta (x_1^0 - x_2^0) \,
   \langle A_1, A_2 | \, \left[ J (x_1) , J (x_2) \right] | A_1, A_2
   \rangle
 \end{align}
 and using the fact that at zero temperature $|G_F|^2 = |G_R|^2$
 \cite{Son:2002sd}, we rewrite \eq{corr6} as
\begin{align}
  \label{corr7}
  \frac{d^6 N_{corr}}{d^2 k_1 \, d y_1 \, d^2 k_2 \, d y_2} \, \propto
  \, \frac{1}{N_c^2} \, |G_R (k_1, k_2)|^2.
\end{align}
Therefore we need to calculate the two-point retarded Green function
at the order $\mu_1 \, \mu_2$. This is exactly the kind of Green
function one can calculate using the AdS/CFT techniques of Eqs.
\peq{Sdiff} and \peq{Gr_mom}.


\section{A Simple Physical Argument}
\label{simple}

Before we present the full calculation of the two-particle
correlations in AdS, we would like to give a simple heuristic argument
of what one may expect from such a calculation. First of all, as we
have noted already, we are going to expand the Green function, and,
therefore, the bulk field $\phi$ into powers of $\mu_1$ and $\mu_2$,
stopping at the order-$\mu_1 \, \mu_2$. To find the field $\phi$ at
the order-$\mu_1 \, \mu_2$ one has to solve \eq{eom} with the metric
taken up to the order $\mu_1 \, \mu_2$. Since we are interested in the
long-range rapidity correlations, our goal is to obtain the leading
rapidity contribution from the calculation. Analyzing Eqs.
\peq{Gr_mom}, \peq{dil_action}, and \peq{Sdiff}, one can conclude that
the leading large-rapidity contribution comes from terms with the
highest number of factors of light-cone momenta, i.e., from terms like
$k_1^+ \, k_2^-$ and $k_1^- \, k_2^+$ (but clearly not from $k_1^+ \,
k_1^- = m_\perp^2/2$ which is rapidity-independent).  Taking $M=N=-$
in \eq{eom} one obtains, among other terms, the following
(leading-rapidity) contribution:
\begin{align}\label{contr1}
g^{--}_{(2)} \ \partial_-^2 \, \phi_0,
\end{align}
where $\phi_0$ is the field at the order $(\mu_1)^0 \, (\mu_2)^0$ and
$g^{MN}_{(2)}$ is the metric at order-$\mu_1 \, \mu_2$.  Concentrating
on order-$z^4$ terms in the metric, which, according to holographic
renormalization \cite{deHaro:2000xn}, are proportional to the
energy-momentum tensor in the boundary theory, and remembering that
the latter is rapidity-independent at order-$\mu_1 \, \mu_2$
\cite{Grumiller:2008va,Albacete:2008vs}, we use energy-momentum
conservation, $\partial_\mu \, T^{\mu\nu} =0$, which, in particular,
implies that $\partial_- \, T^{--} + \partial_+ \, T^{+-} = 0$, to
write
\begin{align}
  g^{--}_{(2)} \, = \, -
  \frac{\partial_+}{\partial_-} \, g^{+-}_{(2)}.
\end{align}
Therefore \eq{contr1} contains the term
\begin{align}\label{contr2}
  - \left( \frac{\partial_+}{\partial_-} \, g^{+-}_{(2)} \right) \
  \partial_-^2 \, \phi_0,
\end{align}
which contributes to the field $\phi$ at order-$\mu_1 \, \mu_2$, and,
as follows from \eq{Gr_mom}, resulting in a contribution to the
retarded Green function in momentum space proportional to
\begin{align}\label{contr3}
  G_R \, \sim \, \frac{k_1^-}{k_1^+} \, {\tilde g}^{+-}_{(2)} \
  (k_2^+)^2
\end{align}
with ${\tilde g}^{+-}$ the Fourier transform of $g^{+-}$ into momentum
space. Since metric component ${\tilde g}^{+-}$ at the order-$\mu_1 \,
\mu_2$ can not be rapidity-dependent
\cite{Grumiller:2008va,Albacete:2008vs}, we see that \eq{contr3} gives
\begin{align}\label{contr4}
  G_R\big|_{|\Delta y| \gg 1} \, \sim \, e^{2 \, (y_2 - y_1)} \, = \,
  e^{2 \, \Delta y}.
\end{align}
Adding the $k_1 \leftrightarrow k_2$ term, arising from the $g^{++}$
component of the metric in \eq{eom}, we get
\begin{align}\label{contr5}
  G_R\big|_{|\Delta y| \gg 1} \, \sim \, \cosh ({2 \, \Delta y}).
\end{align}
Defining the correlation function
\begin{align}\label{corrdef}
  C (k_1, k_2) \, \equiv \, \frac{\frac{d^6 N_{corr}}{d^2 k_1 \, d y_1
      \, d^2 k_2 \, d y_2}}{\frac{d^3 N}{d^2 k_1 \, d y_1} \,
    \frac{d^3 N}{d^2 k_2 \, d y_2}}
\end{align}
and using Eqs. \peq{contr5} and \peq{corr7} to evaluate it we observe
that at large rapidity intervals it scales as
\begin{align}\label{corrf}
  C (k_1, k_2) \big|_{|\Delta y| \gg 1} \, \sim \, \cosh ({4 \, \Delta
    y}).
\end{align}

Indeed the argument we have just presented relies on several
assumptions: in particular it assumes that no other term in the metric
would cancel correlations arising from the terms we have considered.
To make sure that this is indeed the case we will now present the full
calculation.  The result of our simplistic argument given in
\eq{corrf} would still turn out to be valid at the end of this
calculation.


\section{Two-Point Correlation Function at Early Times}
\label{Correlators}


\subsection{Glueball correlator}
\label{glueball}

We now proceed to the calculation of the retarded Green function in
the background of the metric \peq{2nuc_gen}, following the AdS/CFT
prescription outlined in Eqs. \peq{Gr_mom}, \peq{dil_action}, and
\peq{Sdiff}.


\subsubsection{Bulk scalar field}

First we have to find the classical scalar field $\phi$. Similar to
the way the metric \peq{2nuc_gen} was constructed in
\cite{Albacete:2008vs}, we will build the scalar field $\phi$
order-by-order in the powers of $\mu_1$ and $\mu_2$, assuming $\mu_1$
and $\mu_2$ are small perturbations. We would like to find the
solution of \eq{eom} up to order $\cO(\mu_1\mu_2)$. For this we use
the following expansion,
\begin{align}\label{expansion}
  \phi(x,z) = \phi_0(x,z) + \phi_a(x,z) + \phi_b(x,z) + \phi_2(x,z) +
  \ldots \ ,
\end{align}
where $\phi_0 \sim \cO(\mu^0_{1,2})$, $\phi_{a,b} \sim \cO(\mu_{1,2})$
and $\phi_2 \sim \cO(\mu_1\mu_2)$. We will use the standard method
(see e.g. \cite{Mueller:2008bt,Avsar:2009xf,Kovchegov:2010uk}) and
demand that the boundary conditions at $z \rightarrow 0$ are as
follows:
\begin{align}\label{bc}
  \phi_0(x,z \rightarrow 0) \, = \, \phi_B (x), \, \, \, \phi_a(x,z
  \rightarrow 0) \, = \, \phi_b(x,z \rightarrow 0) \, = \, \phi_2(x,z
  \rightarrow 0) \, = \, \ldots \, = \, 0.
\end{align}
In this case the variation of the classical action with respect to
boundary value of the field $\phi_B$ required in \eq{Sdiff} is
straightforward.

Using \eq{2nuc_gen} in \eq{eom}, and expanding the linear operator in
the latter in powers of $\mu_1$ and $\mu_2$ up to order-$\mu_1 \,
\mu_2$ with the help of \peq{LO} and \peq{LOstuff}, the EOM can be
written explicitly in the form
\begin{align}\label{MainEOM}
  \left[ \Box_5 + z^4 \, t_1 \, \partial^2_+ + z^4 \, t_2 \,
    \partial^2_- + \frac{1}{12} \, z^4 \, \hat{M} \right] \, \phi(x,z)
  = 0 \ .
\end{align}
Taking into account that $t_1 = t_1(x^-)$ and $t_2 = t_2(x^+)$, we
give the following list of definitions:
\begin{align}
\Box_5 & \, \equiv \, -\partial^2_z + \frac{3}{z} \, \partial_z + \Box_4 \ ,
%
\ \ \ \ \ \ \ \ \Box_4 \, \equiv \, 2 \, \partial_+\partial_- -
\nabla_{\bot}^2 \ , \ \ \ \ \ \ \ \ \frac{1}{\partial_{\pm}} \equiv
\int^{x^{\pm}}_{-\infty} dx'^{\pm} \ ,
\\ \nonumber %
\hat{M} & \, \equiv \, \left(\hat{D} + z^4 \right) \, t_1 \, t_2 \,
\nabla_{\bot}^2 - \frac{\partial_+}{\partial_-} \, \hat{D} \, t_1 \,
t_2 \, \partial_-^2 - \frac{\partial_-}{\partial_+} \, \hat{D} \, t_1
\, t_2 \, \partial_+^2 + 2 \, \left(\hat{D} + 5 \, z^4 \right) \, t_1
\, t_2 \, \partial_+ \partial_- \\ \nonumber
&+ 5 \, z^4 \, t_1 \, (\partial_+ \, t_2) \, \partial_- + 5 \, z^4 \,
t_2 \, (\partial_- \, t_1) \, \partial_+ + 10 \, z^3 \, t_1 \, t_2 \,
\partial_z + 2 \, z^4 \, t_1 \, t_2 \, \partial_z^2 \ , \\ \nonumber
\hat{D} & \, \equiv \, 96 \, \frac{1}{\partial^2_+} \,
\frac{1}{\partial^2_-} + 16 \, z^2 \, \frac{1}{\partial_+} \,
\frac{1}{\partial_+} + z^4 \ .
\end{align}
Substituting expansion (\ref{expansion}) into (\ref{MainEOM}), and
grouping different powers of $\mu_1$ and $\mu_2$ together we end up
with the following set of equations, listed here along with their
boundary conditions:
\begin{subequations}\label{eom3}
\begin{align}
  &\Box_5 \phi_0(x,z) = 0 \ , \ \ \ \ \ \ \ \ \ \ \phi_0(x,z\to0) = \phi_B(x) \ , \label{eom0} \\
  &\Box_5 \phi_a(x,z) = - z^4 \, t_1(x^-) \, \partial_+^2 \, \phi_0(x,z) \ , \ \ \ \ \ \phi_a(x,z\to0) = 0 \ , \label{eoma} \\
  &\Box_5 \phi_b(x,z) =  - z^4 \, t_2(x^+) \, \partial_-^2 \, \phi_0(x,z) \ , \ \ \ \ \ \phi_b(x,z\to0) = 0 \ , \label{eomb} \\
  &\Box_5 \phi_2(x,z) = - z^4 \, t_1(x^-) \, \partial_+^2 \,
  \phi_b(x,z) - z^4 \, t_2(x^+) \, \partial_-^2 \, \phi_a(x,z) -
  \frac{z^4}{12} \, \hat{M} \, \phi_0(x,z) \ , \ \ \ \ \
  \phi_2(x,z\to0) = 0 \ , \label{eom2}
\end{align}
\end{subequations}
where we also imply that all the solutions should be regular at $z\to
\infty$.

To solve equations \peq{eom3} it is convenient to introduce a Green
function $G(x,z,z')$ satisfying the equation
\begin{align}\label{Green1}
  \Box_5 \, G(x,z,z') \, = \, z'^3 \, \delta(z-z').
\end{align}
The Green function can be written as
\begin{align}\label{Green2}
  G(x,z,z') \, = \, z^2 \, z'^2 \, I_2(z_< \sqrt{\Box_4}) \, K_2(z_>
  \sqrt{\Box_4}) \ ,
\end{align}
where $z_{\{<,>\}} = {\rm \{min,max\}}\{z,z'\}$. We can rewrite the
inverse of $\Box_5$ operator as
\begin{align}
  \frac{1}{\Box_5} f(x,z) \equiv \int^{\infty}_0 \frac{dz'}{z'^3} \,
  G(x,z,z') \, f(x,z').
\end{align}
Solving the first equation in \peq{eom3} we find
\begin{align}\label{free}
  \phi_0(x,z) = \frac{1}{2}z^2 \Box_4 K_2(z\sqrt{\Box_4})\phi_B(x) \ .
\end{align}
From Eqs. \peq{eoma}, \peq{eomb}, and \eq{eom2} we have
\begin{align}
  \phi_a(x,z) &= - \frac{1}{\Box_5} \left[z^4 \, t_1 \, \partial_+^2
    \, \phi_0\right] \ , \ \ \ \ \ \ \ \
  \phi_b(x,z) = - \frac{1}{\Box_5} \left[z^4 \, t_2 \, \partial_-^2 \,
    \phi_0\right] \ , \label{solab} \\
  \phi_2(x,z) &= \frac{1}{\Box_5} \, z^4 \, t_1 \, \partial_+^2 \,
  \frac{1}{\Box_5} \, z^4 \, t_2 \, \partial_-^2 \, \phi_0 +
  \frac{1}{\Box_5} \, z^4 \, t_2 \, \partial_-^2 \, \frac{1}{\Box_5}
  \, z^4 \, t_1 \, \partial_+^2 \, \phi_0 - \frac{1}{\Box_5} \, z^4 \,
  \frac{\hat{M}}{12} \, \phi_0 \ . \label{sol2}
\end{align}
We have constructed the bulk scalar field which we need to find the
correlation function.


\subsubsection{Glueball correlation function}

We can now calculate the retarded glueball correlation function using
\eq{sol2} in Eqs. \peq{dil_action}, \peq{Sdiff}, and \peq{Gr_mom}. It
is straightforward to check that
\begin{align}\label{Gexp}
  &\left[\frac{1}{z^3} \, \partial_z \, G(x,z,z')\right]_{z\to0} \, =
  \, \frac{1}{2} \, z'^2 \, \Box_4 \, K_2(z' \, \sqrt{\Box_4}) \ .
\end{align}
Using \eq{Gexp}, along with Eqs. \peq{dil_action}, \peq{Sdiff}, and
\peq{Gr_mom}, we obtain
\begin{align}
  \label{eq:G1}
  G_R (k_1, k_2) \, = \, \frac{N_c^2}{16} \,\mu_1 \, \mu_2 \,
  \delta^{(2)}({\un k}_{1} + {\un k}_{2}) \, k_1^2 \, k_2^2 \, \left[
    F(k_1,k_2) + F(k_2,k_1) \right] \ ,
\end{align}
where
\begin{align}
  \label{eq:FAB}
  F(k_1,k_2) \equiv & F_\text{I} (k_1,k_2) + F_\text{II} (k_1,k_2)
\end{align}
with
\begin{align}\label{eq:FI}
  F_\text{I} (k_1,k_2) = & \int^{\infty}_0 dz~z^5 \, K_2 \left(z \,
    \sqrt{k_1^2}\right) \,
\int^{\infty}_0 dz'~z'^5 \, K_2\left(z' \, \sqrt{k_2^2} \right) \notag \\
& \times \, \left[(k_1^-k_2^+)^2
  I_2\left(Q_1z_<\right)K_2\left(Q_1z_>\right) + (k_1^+k_2^-)^2
  I_2\left(Q_2z_<\right)K_2\left(Q_2z_>\right)\right]
\end{align}
and
\begin{align}\label{eq:FII}
  F_\text{II} (k_1,k_2) = &\frac{k^2_{2\bot}}{12}\int^{\infty}_0 dz~z^5 K_2
  \left(z \, \sqrt{k_1^2} \right)
\left[\frac{96}{(k_1^+k_1^-)^2} - \frac{16 \, z^2}{k_1^+k_1^-}\right]
K_2\left(z \, \sqrt{k_2^2}\right) \notag \\[7pt] \nonumber
&-
\frac{1}{12}\left[\frac{k_1^-k_2^{+2}}{k_1^+}+\frac{k_1^+k_2^{-2}}{k_1^-}\right] \,
\int^{\infty}_0 dz~z^5 K_2 \left( z \, \sqrt{k_1^2} \right)
\left[\frac{96}{(k_1^+k_1^-)^2} - \frac{16 \, z^2}{k_1^+k_1^-} + z^4 \right]
\, K_2 \left(z \, \sqrt{k_2^2} \right) \\[7pt] \nonumber
&+\frac{1}{6} \, k_2^+k_2^-\int^{\infty}_0 dz~z^5 \, K_2 \left( z \,
  \sqrt{k_1^2} \right)
\left[\frac{96}{(k_1^+k_1^-)^2} - \frac{16 \, z^2}{k_1^+k_1^-} + 8 \,
  z^4 \right] K_2 \left(z \, \sqrt{k_2^2} \right) \\[7pt] \nonumber &-
\frac{5}{12}\left[2 \, k_2^+k_2^- + k_2^+k_1^- +
  k_2^-k_1^+\right]\int^{\infty}_0 dz~z^9 \, K_2 \left( z\,
  \sqrt{k_1^2} \right)
K_2 \left(z \, \sqrt{k_2^2} \right) \\[7pt]
&+ \frac{4 \, k_2^2}{3} \, \int^{\infty}_0 dz~z^8 \, K_2 \left( z\,
  \sqrt{k_1^2} \right) \, K_1\left(z \, \sqrt{k_2^2} \right)\ .
\end{align}
We have defined
\begin{align}
  \label{eq:Q12}
  Q_1^2 \, = \, 2 \, k_1^- \, k_2^+ + k_\perp^2, \ \ \ Q_2^2 \, = \, 2
  \, k_1^+ \, k_2^- + k_\perp^2,
\end{align}
with $k_{1, \perp} = k_{2, \perp} = k_\perp$.

Before evaluating the obtained expressions further, let us comment on
some of their features. First one may note that \eq{eq:G1}
contains a delta-function of transverse momenta of the two glueballs
$\delta^{(2)}({\un k}_{1} + {\un k}_{2})$. This demonstrates that at
the lowest non-trivial order in $\mu_1$ and $\mu_2$ expansion
(order-$\mu_1 \, \mu_2$) there will be nothing else produced in the
shock wave collision apart from the two glueballs. Note that indeed a
non-zero $\langle T_{\mu\nu} \rangle$ in the forward light-cone at the
order-$\mu_1 \, \mu_2$ found in
\cite{Grumiller:2008va,Albacete:2008vs} indicates that a medium is
created: however this strongly-coupled medium in the ${\cal N} =4$ SYM
theory without bound states and confinement does not fragment into
individual particles, and at late times simply results in a very low
(and decreasing) energy density created in the collision, similar to
the asymptotic future of Bjorken hydrodynamics dual found in
\cite{Janik:2005zt}. Since in our calculation we have explicitly
projected out two glueballs with fixed momenta in the final state,
those two glueballs are all that is left carrying transverse momentum
in the forward light-cone. (Leftovers of the original shock waves may
also be present, though they would not carry any transverse momentum.)
This picture is in agreement with the dominance of elastic processes
in high energy scattering in the AdS/CFT framework suggested in
\cite{Levin:2008vj}.

Another important aspect of the result in Eqs. \peq{eq:FI} and
\peq{eq:FII} above is that the integrals over $z$ and $z'$ diverge for
time-like momenta $k_1$ and $k_2$, i.e., for $k_1^2 = - m^2$ and
$k_2^2 = - m^2$ corresponding to production of physical glueballs of
mass $m$. This result should be expected in ${\cal N} =4$ SYM theory:
since there are no bound states in this theory, we conclude that there
are no glueballs. Thinking of Bessel functions $K_2 (z
\sqrt{k_{1,2}^2})$ in Eqs. \peq{eq:FI} and \peq{eq:FII} as
contributing to the wave functions of glueballs in AdS$_5$ space
\cite{Brodsky:2003px,Polchinski:2002jw,Polchinski:2000uf}, we conclude
that the lack of glueball bound states in the theory manifests itself
through de-localization of these wave functions, resulting in ``bound
states'' of infinite radii, both in the bulk and in the boundary
theory (if we identify the holographic coordinate $z$ with the inverse
momentum scale on the UV boundary). Since the glueballs for us have
always been some external probes of the ${\cal N} =4$ SYM theory, we
conclude that one has to define the probes by re-defining their
wavefunctions. This can be accomplished, for instance, by introducing
confinement in the theory, by using either the ``hard-wall'' or
``soft-wall'' models
\cite{Polchinski:2001tt,BoschiFilho:2002vd,Brodsky:2003px,Erlich:2005qh,DaRold:2005zs,BoschiFilho:2005yh,Grigoryan:2007vg,Karch:2006pv,Karch:2010eg,Grigoryan:2007my}.
The inverse confinement scale would define the typical size of the
bound states. Indeed such procedure would introduce a model-dependent
uncertainty associated with mimicking confinement in AdS/CFT, but is
unavoidable in order to define glueball probes. Besides, our main goal
here is to calculate long-range rapidity correlations, which are not
affected (apart from a prefactor) by the exact shape of the glueball
AdS$_5$ wave functions.  We therefore model confinement by modeling
the glueball (external source) AdS wave functions by simply replacing
$K_2 (z \sqrt{k_{1,2}^2}) \rightarrow K_2 (z \, \Lambda)$ in Eqs.
\peq{eq:FI} and \peq{eq:FII} with $\Lambda >0$ related to confinement
momentum scale. We then rewrite Eqs. \peq{eq:FI} and \peq{eq:FII} as
\begin{align}\label{eq:FI2}
  F_\text{I} (k_1,k_2) = & \int^{\infty}_0 dz~z^5 \, K_2 \left(z \,
    \Lambda \right) \,
  \int^{\infty}_0 dz'~z'^5 \, K_2\left(z' \, \Lambda \right) \notag \\
  & \times \, \left[(k_1^-k_2^+)^2
    I_2\left(Q_1z_<\right)K_2\left(Q_1z_>\right) + (k_1^+k_2^-)^2
    I_2\left(Q_2z_<\right)K_2\left(Q_2z_>\right)\right],
\end{align}
\begin{align}\label{eq:FII2}
  F_\text{II} (k_1,k_2) = &\frac{k^2_{\bot}}{12}\int^{\infty}_0 dz~z^5
  \, \left[ K_2 \left(z \, \Lambda \right) \right]^2
  \left[\frac{384}{m_\perp^4} - \frac{32 \, z^2}{m_\perp^2}\right]
  \notag \\[7pt] \nonumber
  &-
  \frac{1}{12}\left[\frac{k_1^-k_2^{+2}}{k_1^+}+\frac{k_1^+k_2^{-2}}{k_1^-}\right]
  \, \int^{\infty}_0 dz~z^5 \, \left[ K_2 \left( z \, \Lambda \right)
  \right]^2
\left[\frac{384}{m_\perp^4} - \frac{32 \, z^2}{m_\perp^2} + z^4
\right] \\[7pt] \nonumber
&+\frac{1}{12} \, m_\perp^2 \, \int^{\infty}_0 dz~z^5 \, \left[ K_2
  \left( z \, \Lambda \right) \right]^2
\left[\frac{384}{m_\perp^4} - \frac{32 \, z^2}{m_\perp^2} + 8 \, z^4
\right] \\[7pt] \nonumber &- \frac{5}{12}\left[m_\perp^2 + k_2^+k_1^-
  + k_2^-k_1^+\right]\int^{\infty}_0 dz~z^9 \, \left[ K_2 \left( z\,
    \Lambda \right) \right]^2 \\[7pt]
&+ \frac{4 \, m^2}{3} \, \int^{\infty}_0 dz~z^8 \, K_2 \left( z\,
  \Lambda \right) \, K_1\left(z \, \Lambda \right)\ ,
\end{align}
where we have also replaced all rapidity-independent factors with
powers of either glueball mass $m$ or $m_\perp = \sqrt{k_\perp^2 +
  m^2}$.

\begin{figure}[th]
\begin{center}
\epsfxsize=9cm
\leavevmode
\hbox{\epsffile{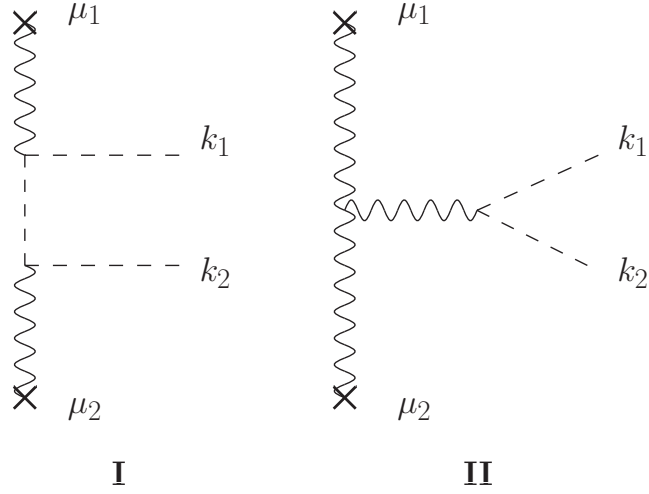}}
\end{center}
\caption{Diagrammatic representation of the correlation function
     calculated in this Section.}
\label{graphs}
\end{figure}

The contributions in Eqs. \peq{eq:FI2} and \peq{eq:FII2} (or those in
Eqs. \peq{eq:FI} and \peq{eq:FII}) to the retarded Green function
\peq{Gr_mom} are shown diagrammatically in \fig{graphs} in terms of
Witten diagrams.  There the wiggly lines represent gravitons, while
the dashed line denotes the scalar field. Crosses represent insertions
of the boundary energy-momentum tensors of the two shock waves
($\mu_1$ and $\mu_2$).  $F_\text{I}$ from \eq{eq:FI2} corresponds to
the diagram on the left of \fig{graphs}, while $F_\text{II}$ from
\eq{eq:FII2} is given by the term on the right of \fig{graphs}.

It is important to note that the Green function given by Eqs.
\peq{eq:G1}, \peq{eq:FAB}, \peq{eq:FI2}, and \peq{eq:FII2} is indeed
real, justifying the assumption we employed in stating that \eq{Sdiff}
provides us a retarded Green function. This can also be seen from the
diagrams in \fig{graphs} in which one can not cut the scalar
propagator. The imaginary part of $G_R$ appears at higher order in
$\mu_1 \, \mu_2$, when one has more graviton insertions in the scalar
propagator, allowing for non-zero cuts of the latter.

Let us now study the large-rapidity interval asymptotics of the
obtained correlation function \peq{eq:G1}.  One can deduce from the
kinematics described in Section \ref{kine} that
\begin{align}
  &k^+_1k^-_2 = \frac{m^2_{\bot}}{2} \, e^{-\Delta y} \ , \ \ \ \ \
  k^-_1k^+_2 = \frac{m^2_{\bot}}{2} \, e^{\Delta y} \ ,
\end{align}
such that when $\Delta y = y_2 - y_1 \gg 1$ we have
\begin{align}
  Q^2_1 = k^2_{\bot} + m_{\bot}^2 \, e^{\Delta y} \approx m^2_{\bot}
  \, e^{\Delta y} \ , \ \ \ Q^2_2 = k^2_{\bot} + 2 \, m_{\bot}^2 \,
  e^{-\Delta y} \approx k^2_{\bot}.
\end{align}
Therefore, the contribution from \eq{eq:FI2} becomes
\begin{align}\label{FIapp1}
  F_I (k_1,k_2) \big|_{\Delta y \gg 1} \, \approx \, \int^{\infty}_0
  dz~z^5 \, K_2 \left(z \, \Lambda \right) \, \int^{\infty}_0 dz'~z'^5
  \, K_2\left(z' \, \Lambda \right) \, (k_1^-k_2^+)^2
  I_2\left(Q_1z_<\right) \, K_2 \left(Q_1z_>\right).
\end{align}
To determine the large-$Q_1$ asymptotics of $I_2\left(Q_1z_<\right) \,
K_2 \left(Q_1z_>\right)$ note that, according to Eqs. \peq{Green1} and
\peq{Green2}, $z^2 \, z'^2 \, I_2\left(Q_1z_<\right) \, K_2
\left(Q_1z_>\right)$ satisfies
\begin{align}
  \label{eq:IK}
  \left[ - \partial_z^2 + \frac{3}{z} \, \partial_z + Q_1^2 \right] \,
  z^2 \, z'^2 \, I_2\left(Q_1z_<\right) \, K_2 \left(Q_1z_>\right) \,
  = \, z'^3 \, \delta (z - z').
\end{align}
Hence, for $Q_1$ larger than the inverse of the typical variation in
$z$ we have
\begin{align}
  \label{eq:IK2}
  z^2 \, z'^2 \, I_2\left(Q_1z_<\right) \, K_2 \left(Q_1z_>\right)
  \bigg|_{\text{large} \, Q_1} \, \approx \, \frac{z'^3}{Q_1^2} \, \delta
  (z - z'),
\end{align}
which, when used in \eq{FIapp1} yields
\begin{align}
  \label{FIapp2}
  F_I (k_1,k_2) \big|_{\Delta y \gg 1} \, \approx \, \frac{2048}{7} \,
  \frac{(k_1^-k_2^+)^2}{Q_1^2 \, \Lambda^{10}} \, \approx \,
  \frac{512}{7} \, \frac{m_\perp^2}{\Lambda^{10}} \, e^{\Delta y}.
\end{align}
This result implies that the rapidity correlations coming from this
term grow as $e^{\Delta y}$ at the early stages after the collision.

On the other hand, the dominant contributions from the second term,
$F_\text{II} (k_1,k_2)$, are coming from the expressions in the second
and the fourth lines of \eq{eq:FII2}. They give
\begin{align}\label{FIIapp}
  F_\text{II} (k_1,k_2) \big|_{\Delta y \gg 1} &\approx -
  \frac{1}{12}\left[\frac{k_1^-k_2^{+2}}{k_1^+}+\frac{k_1^+k_2^{-2}}{k_1^-}\right]
  \, \int^{\infty}_0 dz~z^5 \, \left[ K_2 \left( z \, \Lambda \right)
  \right]^2 \left[\frac{384}{m_\perp^4} - \frac{32 \, z^2}{m_\perp^2}
    + z^4 \right] \\ \nonumber &- \frac{5}{12} \, \left(k_2^+k_1^- +
    k_1^+k_2^-\right) \, \int^{\infty}_0 dz~z^9 \, \left[ K_2 \left( z
      \, \Lambda \right) \right]^2 \\ \nonumber & \approx -
  \frac{256}{21} \, \frac{m^2_{\bot}}{\Lambda^{10}} \, e^{2 \, \Delta
    y} \, \left[ 1 - 3 \, \frac{\Lambda^2}{m_\perp^2} + \frac{42}{5}
    \, \frac{\Lambda^4}{m_\perp^4} \right] - \frac{1280}{21} \,
  \frac{m^2_{\bot}}{\Lambda^{10}} \, e^{\Delta y}.
\end{align}

Combining Eqs. \peq{FIapp2} and \peq{FIIapp} in Eqs. \peq{eq:G1} and
\peq{eq:FAB} we obtain
\begin{align}
  \label{GRlargey}
  G_R (k_1, k_2)\big|_{\Delta y \gg 1} \, \approx \, - \frac{64}{21}
  \, \frac{N_c^2 \,\mu_1 \, \mu_2 \, m^4 \, m_\perp^2}{\Lambda^{10}}
  \, \delta^{(2)}({\un k}_{1} + {\un k}_{2}) \, \left\{ e^{2 \, \Delta
      y} \, \left[ 1 - 3 \, \frac{\Lambda^2}{m_\perp^2} + \frac{42}{5}
      \, \frac{\Lambda^4}{m_\perp^4} \right] + e^{\Delta y} \right\},
\end{align}
which, dropping the second term in the curly brackets and using the $+
\leftrightarrow -$ symmetry of the problem can be generalized to
\begin{align}
  \label{GR_gen}
  G_R (k_1, k_2)\big|_{|\Delta y| \gg 1} \, \approx \, & -
  \frac{128}{21} \, \frac{N_c^2 \,\mu_1 \, \mu_2 \, m^4 \,
    m_\perp^2}{\Lambda^{10}} \, \delta^{(2)}({\un k}_{1} + {\un
    k}_{2}) \, \cosh ({2 \, \Delta y}) \, \left[ 1 - 3 \,
    \frac{\Lambda^2}{m_\perp^2} + \frac{42}{5} \,
    \frac{\Lambda^4}{m_\perp^4} \right].
\end{align}
We thus conclude that at large rapidity separations
\begin{align}
  \label{GRlarge}
  G_R (k_1, k_2)\big|_{|\Delta y| \gg 1} \, \sim \, \cosh ({2 \,
    \Delta y})
\end{align}
in agreement with our estimate in \eq{contr5}.

Using \eq{corr7} we conclude that
\begin{align}
  \frac{d^6 N_{corr}}{d^2 k_1 \, d y_1 \, d^2 k_2 \, d y_2}
  \Bigg|_{|\Delta y| \gg 1} \, \sim \, \cosh ({4 \, \Delta y})
\end{align}
such that the two-glueball correlation function defined in
\eq{corrdef} scales as
\begin{align}\label{Corr_fin}
  C (k_1, k_2) \big|_{|\Delta y| \gg 1} \, \sim \, \cosh ({4 \, \Delta
    y}),
\end{align}
just like in \eq{corrf}. We have demonstrated the presence of
long-range rapidity correlations in case of strongly-coupled
high-energy heavy ion collisions. The rapidity shape of the obtained
correlations is very different from the ``ridge'' correlation observed
experimentally at RHIC and at LHC
\cite{Adams:2005ph,Adare:2008cqb,Alver:2009id,Khachatryan:2010gv}. It
is possible that higher order in $\mu_1$ and $\mu_2$ corrections would
modify the rapidity shape of the correlation, putting it more in-line
with experiments. We will return to this point in Sec. \ref{sum}.

Let us now pause to determine the parameter of our approximation.
Until now we have, somewhat loosely, referred to our approximation as
to an expansion in $\mu_1$ and $\mu_2$. However, these parameters have
dimensions of mass cubed, and can not be expanded in. From \eq{GR_gen}
we may suggest that the dimensionless expansion parameters are $\mu_1
/ \Lambda^3$ and $\mu_2 / \Lambda^3$, where $\Lambda$ is the inverse
glueball size. Thus our result in \eq{GR_gen} dominates the
correlation function only for
\begin{align}
  \label{conds}
  \frac{\mu_1}{\Lambda^3} \ll 1, \ \ \ \frac{\mu_2}{\Lambda^3} \ll 1.
\end{align}
Since, as can be seen from \eq{mus}, $\mu_1$ and $\mu_2$ are
energy-dependent, these conditions limit the energy range of
applicability of \eq{GR_gen}. \eq{conds} also makes clear physical
sense: since the metric \peq{2nuc_gen} with the coefficients given by
Eqs. \peq{LO} and \eq{LOstuff} is valid only for early proper times
$\tau$ satisfying $\mu_{1,2} \, \tau^3 \ll 1$
\cite{Albacete:2008vs,Albacete:2009ji}, we see that the glueballs have
to be small enough, $1/\Lambda \approx \tau \approx \mu_{1,2}^{-1/3}$,
to be able to resolve (and be sensitive to) the metric at such early
times.

Note also that the obtained Green function \peq{GR_gen} is not a
monotonic function of $m_\perp$: for $m_\perp \ll \Lambda$ it grows
with $m_\perp$ as $m_\perp^2$, but, for $m_\perp \gg \Lambda$ it falls
off as $1/m_\perp^2$, peaking at $m_\perp^2 = (28/5) \, \Lambda^2$.
This translates into correlation function $C (k_1, k_2)$ first growing
with $m_\perp$ (and, therefore, $k_\perp$) as $m_\perp^4$ for $m_\perp
\ll \Lambda$, and then decreasing as $1/m_\perp^4$ for $m_\perp \ll
\Lambda$.  Similar non-monotonic behavior has been observed for
``ridge'' correlation experimentally
\cite{Adams:2005ph,Adare:2008cqb,Alver:2009id,Khachatryan:2010gv}.
While in CGC-based approaches
\cite{Dumitru:2008wn,Gavin:2008ev,Dusling:2009ni,Dumitru:2010iy,Dumitru:2010mv,Kovner:2010xk}
the maximum of the correlation function is given by the saturation
scale $Q_s$, and happens at $k_\perp \approx Q_s$, in our AdS/CFT case
the maximum appears to be related to the inverse size of the produced
bound state and its mass, such that it takes place at $k_\perp \approx
\sqrt{\Lambda^2 - m^2}$. At this point it is not clear though whether
such conclusion is a physical prediction or an artifact of the
perturbative solution of the problem in the AdS space.

In order to make a more detailed comparison with experiment one needs
to improve on our AdS/CFT approach both by calculating higher-order
corrections in $\mu_1$ and $\mu_2$, and, possibly, by implementing
non-conformal QCD features, such as confinement, along the lines of
the AdS/QCD models
\cite{Polchinski:2001tt,BoschiFilho:2002vd,Brodsky:2003px,Erlich:2005qh,DaRold:2005zs,BoschiFilho:2005yh,Grigoryan:2007vg,Karch:2006pv,Karch:2010eg,Grigoryan:2007my}.
The latter modification would certainly change our glueball wave
functions in the bulk, modifying the Bessel functions in
Eqs.~\peq{eq:FI2} and \peq{eq:FII2}. However, while the use of AdS/QCD
geometry may affect the $m_\perp$-dependence of the correlation
function \peq{GR_gen}, one may see from Eqs.~\peq{eq:FI2} and
\peq{eq:FII2} that such modification would not affect our main
conclusion about the rapidity-dependence of the correlations shown in
\eq{Corr_fin}. The leading large-rapidity asymptotics of the
correlation function \peq{Corr_fin} results from the second term on
the right-hand-side of \eq{eq:FII2}: modifying the glueball wave
function would only change the coefficient in front of the
rapidity-dependent part.\footnote{As the integrand in that term is
  positive-definite for any glueball wave function, the coefficient
  can not vanish.} Since the growth of correlations with rapidity does
not reproduce experimental data
\cite{Adams:2005ph,Adare:2008cqb,Alver:2009id,Khachatryan:2010gv}, our
conclusion is that the inclusion of higher-order corrections in
$\mu_1$ and $\mu_2$ is the only possibility for AdS/CFT (or AdS/QCD)
calculations to get in line with the data.


\subsection{Energy-momentum tensor correlator}

We have shown that there are long-range rapidity correlations in the
glueball operator of \eq{Jdef} in the strong-coupling heavy ion
collisions. At the same time we would like to extend this statement to
correlations of other operators. Energy-momentum tensor is a natural
next candidate. Indeed the glueball operator \peq{Jdef} is a part of
the energy-momentum tensor: hence correlations in $\langle J(x) \,
J(y) \rangle$ probably imply correlations in $\langle T_{\mu\nu} (x)
\, T_{\mu\nu} (y) \rangle$ as well. To show this is true we will
present an argument below, largely following
\cite{Policastro:2002se,Kovtun:2004de}.

Consider a field theory whose dual holographic description is given
by the metric of the general form
\begin{align}\label{genmet}
  ds^2 &= g^{(0)}_{MN} \, dx^M \, dx^N = f(x^+,x^-,z) \, dx^2_{\bot} +
  g_{\mu\nu}(x^+,x^-,z) \, d\xi^{\mu} \, d\xi^{\nu} \ ,
\end{align}
where ${\un x} = (x^1, x^2)$,
$d x_\perp^2 = (d x^1)^2 + (d x^2)^2$,
and $\xi^{\mu} = (x^+,x^-,z)$. Now, consider small perturbations
around the metric independent of $x^1, x^2$, $g_{MN}= g_{MN}^{(0)} +
h_{MN} (x^+,x^-,z)$. We will work in the $h_{Mz} =0$ gauge. The metric
\peq{genmet} has a rotational $O$(2) symmetry in the transverse plane.
Under the transverse rotations one may naively expect $\{ h_{11},
h_{12}, h_{22} \}$ components to transform as tensors, $\{ h_{01},
h_{31}, h_{02}, h_{32} \}$ components to transform as vectors, and $\{
h_{00}, h_{03}, h_{33} \}$ components to be scalars under rotations.
However, rewriting the transverse part of the metric as
\begin{align}
  \label{trmet}
  \left(
    \begin{array}{cc}
        h_{11} & h_{12} \\
        h_{21} & h_{22}
    \end{array}
\right) \, = \,
\left(
    \begin{array}{cc}
        (h_{11} + h_{22})/2 & 0 \\
        0 & (h_{11} + h_{22})/2
    \end{array}
\right) +
\left(
    \begin{array}{cc}
        (h_{11} - h_{22})/2 & h_{12} \\
        h_{21} & -  (h_{11} - h_{22})/2
    \end{array}
\right)
\end{align}
we see that $h_{11} + h_{22}$ is also invariant under $O$(2)
transverse plane rotations. Hence the final classification of the
metric components under $O$(2) rotations is: $\{ h_{11} - h_{22},
h_{12} \}$ are in the tensor representation, $\{ h_{01}, h_{31},
h_{02}, h_{32} \}$ are vectors, and $\{ h_{00}, h_{03}, h_{33}, h_{11}
+ h_{22} \}$ are scalars \cite{Policastro:2002se,Kovtun:2004de}.

Using the above classification we see that we can assume that the only
non-vanishing component of $h_{MN}$ is $h_{12} = h_{21} =
h_{12}(x^+,x^-,z)$. It is in the tensor representation and, as can be
seen with the help of \eq{trmet}, by rotating in the transverse plane
we can always find a coordinate system in which $h_{11} - h_{22} =0$
and $h_{12} = h_{21}$ remains the only non-zero metric component in
the tensor representation. Since all other components of the metric
are in other representations of the $O$(2) symmetry group, they do not
mix with $h_{12}$ in Einstein equations, and can be safely put to zero
\cite{Policastro:2002se,Kovtun:2004de}.

Substituting the metric $g_{MN}= g_{MN}^{(0)} + h_{MN} (x^+,x^-,z)$
with $g_{MN}^{(0)}$ given by \peq{genmet} into Einstein equations
\peq{ein}, and expanding the result to linear order in $h_{12}$ we get
\cite{Policastro:2002se,Kovtun:2004de}
\begin{align}
  \Box \, h_{12} - 2 \, \frac{\partial^{\mu} f}{f} \, \partial_{\mu}
  h_{12} + 2 \, \frac{(\partial f)^2}{f^2} \, h_{12} - \frac{\Box
    f}{f} \, h_{12} = 0\, ,
\label{minscal}
\end{align}
where
\begin{align}
  \label{box3}
  \Box \, = \, \frac{1}{\sqrt{-g}} \, \partial_M \left[ \sqrt{-g} \, g^{MN} \, \partial_N \ldots \right] 
\end{align}
and $(\partial f)^2 = g^{MN} \, \partial_M f \, \partial_N f$.
Changing the variable from $h_{12}$ to $h^1_2=h_{12}/f$, one can see
that $h^1_2$ indeed satisfies the equation for a minimally coupled
massless scalar \cite{Policastro:2002se,Kovtun:2004de}:
\begin{align}
  \Box \, h^1_2 = 0.
\end{align}
Therefore, since our metric \peq{2nuc_gen} falls into the category of
\eq{genmet}, the analysis of Sec \ref{glueball} applies to the metric
component $h_2^1$. Defining the retarded Green function for the
$T_2^1$ components of the energy-momentum tensor (EMT) by
\begin{align}
  \label{Gr_ten}
  G^{EMT}_R (k_1, k_2) \, = \, - i \, \int d^4 x_1 \, d^4 x_2 \, e^{-
    i \, k_1 \cdot x_1 - i \, k_2 \cdot x_2} \, \theta (x_1^0 - x_2^0)
  \, \langle A_1, A_2 | \, \left[ T_2^1 (x_1) , T_2^1 (x_2) \right] |
  A_1, A_2 \rangle
\end{align}
we conclude that, similar to the glueball operator,
\begin{align}
  \label{GRlargeEMT}
  G^{EMT}_R (k_1, k_2)\big|_{|\Delta y| \gg 1} \, \sim \, \cosh ({2 \,
    \Delta y}).
\end{align}
Hence we have shown that the correlators of EMT operators exhibit the
same long-range rapidity correlations as the glueball correlators. It
is therefore very likely that such correlations are universal and are
also present in correlators of other operators.


\section{Estimate of the Two-Point Correlation Function at Late Times}
\label{late-times}

Our conclusion about long-range rapidity correlations was derived
using the metric \peq{2nuc_gen} which is valid only at very early
times after a shock wave collision. As discussed in the Introduction,
we do not expect the interactions at later times to affect these
correlations, since different-rapidity regions of the produced medium
become causally disconnected at late times. To check that no
long-range rapidity correlations can arise from the late-time dynamics
one would have to calculate the correlation function \peq{corrdef} in
the full metric produced in a shock wave collision including all
powers of $\mu_1$ and $\mu_2$. Since no such analytical solution
exists, instead we will use the metric dual to Bjorken hydrodynamics
\cite{Bjorken:1982qr} constructed in \cite{Janik:2005zt}. One has to
be careful in interpreting the result we obtain in this Section:
Bjorken hydrodynamics \cite{Bjorken:1982qr} is rapidity-independent,
while there are reasons to believe that the medium produced in a shock
wave collision would exhibit rapidity dependence, as indicated by
perturbative solutions of Einstein equations done in
\cite{Grumiller:2008va,Albacete:2008vs,Albacete:2009ji}. Nonetheless,
we expect that our calculation below would be a good initial estimate
of the late-time rapidity correlations.

The dual geometry corresponding to the perfect fluid was obtained by
Janik and Peschanski in \cite{Janik:2005zt}. It can be written as
\begin{align}\label{JPmetric}
  ds^2 = L^2 \, \left\{ -\frac{1}{z^2}\frac{\left(1 -
        z^4/z_h^4(\tau)\right)^2}{1 + z^4/z_h^4(\tau)}d\tau^2 +
    \frac{\left(1 + z^4/z_h^4(\tau)\right)}{z^2}\left(\tau^2 d\eta^2 +
      dx_{\bot}^2\right) + \frac{dz^2}{z^2} \right\} \ ,
\end{align}
where $\tau = \sqrt{2x^+x^-}$ is proper time, $\eta =
\frac{1}{2}\ln(x^+/x^-)$ is space-time rapidity, and $z_h(\tau) =
\left(\frac{3}{\cE_0}\right)^{1/4}\tau^{1/3}$ (with $\cE_0$ some
dimensionful quantity) determines the position of the dynamical
horizon in AdS$_5$ such that the Hawking temperature is
\begin{align}
T(\tau) = \frac{\sqrt{2}}{\pi z_h(\tau)} =
\frac{\sqrt{2}}{\pi}\left(\frac{\cE_0}{3}\right)^{1/4}~\tau^{-1/3} \
.
\end{align}

Unfortunately finding the glueball correlation function in Bjorken
hydrodynamic state is equivalent to finding boundary-to-boundary
scalar propagator in the background of the Janik-Peschanski metric
\peq{JPmetric}, which is a daunting task: such propagator has not yet
been found even for the static AdS Schwarzschild black hole metric.
Instead, to estimate the correlations we will perform a perturbative
calculation.

At late times, when $\tau \gg \cE_0^{-3/8}$, assuming either that $z$
is fixed or is bounded from the above (by let us say an infrared (IR)
cutoff coming from the definition of the glueball wave function), we
can consider the ratio $u(\tau) \equiv z/z_{h}(\tau) \ll 1 $ to be a
small quantity. If so, we can expand the EOM for the scalar field
\peq{eom} up to $\cO(u^4)$ obtaining
\begin{align}\label{EOMJP}
  &\Box_5 \phi(\tau, \eta, x_{\bot}, z) + u^4 \, \left[4 \,
    \partial^2_{\tau}
    - \Box_4 \right] \, \phi(\tau, \eta, x_{\bot}, z) = 0 \ , \\
  \nonumber
&\Box_5\phi \equiv -z^3 \partial_z\left(\frac{1}{z^3}\partial_z
\phi\right) + \Box_4 \phi \ , \ \ \ \ \ \
\Box_4\phi \equiv \frac{1}{\tau}\partial_{\tau}\left(\tau
  \partial_{\tau}\phi\right) - \frac{1}{\tau^2}\partial^2_{\eta}\phi -
\nabla^2_{\bot}\phi = \left(2\partial_+\partial_- -
  \nabla^2_{\bot}\right)\phi \ .
\end{align}

Expanding the scalar field in the powers of $u$ we write
\begin{align}
  \label{eq:phi_exp}
  \phi = \phi_0 + \phi_1 + \ldots
\end{align}
where $\phi_0 \sim \cO\left(u^0 \right)$ and $\phi_1 \sim
\cO\left(u^{4}\right)$. Substituting this back into Eq.~(\ref{EOMJP}),
we get
\begin{align}\label{EOMJP2}
  &\Box_5 \, \phi_0 = 0 \ , \ \ \ \ \ \ \ \Box_5 \, \phi_1 = -
  \frac{\cE_0}{3} \, \frac{z^4}{\tau^{4/3}} \, \left[4 \,
    \partial^2_{\tau} - \Box_4\right] \, \phi_0 \ .
\end{align}
The solution for $\phi_0$ was found above and is given in \eq{free}.
We write the solution for $\phi_1$ as
\begin{align}
  \label{phi1}
  \phi_1 = - \frac{\cE_0}{3} \, \frac{1}{\Box_5} \,
  \frac{z^4}{\tau^{4/3}} \, \left[4 \, \partial^2_{\tau} - \Box_4
  \right] \, \phi_0 \, \approx \, \frac{\cE_0}{3} \, \frac{1}{\Box_5}
  \, \frac{z^4}{\tau^{4/3}} \, \Box_4 \, \phi_0
\end{align}
where in the last step we neglected $\partial^2_{\tau}$, since a
derivative like this generates $O (1/\tau^2)$ corrections (at fixed
$u$), which were neglected in constructing the original metric
\peq{JPmetric} and are thus outside of the precision of our
approximation.  We are now ready to calculate the retarded Green
function. Using \eq{phi1} in Eqs. \peq{Sdiff}, \peq{dil_action}, and
\peq{Gr_mom}, and employing \eq{Gexp} yields
\begin{align}
  \label{GrBj1}
  G_R^{Bj} (k_1, k_2)\big|_{O(1/z_h^4)} \, = \, - \frac{N_c^2 \, \cE_0
    \, m^6}{24} \, \delta^2 ({\un k}_1 + {\un k}_2) \, \int^{\infty}_0
  dz~z^5 \, K_2 \left( z \, \sqrt{k_1^2} \right) \, K_2 \left( z \,
    \sqrt{k_2^2} \right) \notag \\ \times \, \int_0^\infty d x^+ \, d
  x^- \, e^{i \, x^+ \, (k_1^- + k_2^-) + i \, x^- \, (k_1^+ + k_2^+)}
  \, \frac{1}{\tau^{4/3}}
\end{align}
where we have replaced $k_1^2$ and $k_2^2$ with $-m^2$ everywhere
except for the arguments of the Bessel functions.  The integrals over
$x^+$ and $x^-$ in \eq{GrBj1} run from $0$ to $\infty$ since the
matter only exists in the forward light-cone. (On top of that the
metric \peq{JPmetric} is valid at late times only, for $u \ll 1$, such
that the actual $x^+$ and $x^-$ integration region should be even more
restricted, possibly suppressing the correlations we are about to
obtain even more.)

Just like in the case of the early times considered in Sec.
\ref{glueball}, the integral over $z$ in \eq{GrBj1} is divergent for
time-like momenta $k_1$ and $k_2$. Similar to what we did in Sec.
\ref{glueball}, we recognize the Bessel functions in \eq{GrBj1} as the
glueball wave functions in the bulk, which need to be modified to
reflect the finite size of glueballs, which do not exist in ${\cal N}
=4$ SYM theory. Replacing $K_2 (z \sqrt{k_{1,2}^2}) \rightarrow K_2 (z
\, \Lambda)$ in \eq{GrBj1} and integrating over $z$ yields
\begin{align}
  \label{GrBj2}
  G_R^{Bj} (k_1, k_2)\big|_{O(1/z_h^4)} \, = \, - \frac{4 \, N_c^2 \,
    \cE_0 \, m^6}{15 \, \Lambda^6} \, \delta^2 ({\un k}_1 + {\un k}_2)
  \, \int_0^\infty d x^+ \, d x^- \, e^{i \, x^+ \, (k_1^- + k_2^-) +
    i \, x^- \, (k_1^+ + k_2^+)} \, \frac{1}{\tau^{4/3}}.
\end{align}

Evaluating the integrals left in \eq{GrBj2},
\begin{align}
  \int_0^\infty d x^+ \, d x^- \, e^{i \, x^+ \, (k_1^- + k_2^-) + i
    \, x^- \, (k_1^+ + k_2^+)} \, \frac{1}{(2 \, x^+ \, x^-)^{2/3}} \,
  = \, \frac{N}{(k_1^+ + k_2^+)^{1/3}(k_1^- + k_2^-)^{1/3}} \ ,
\end{align}
where
\begin{align}
  N \, = \, \frac{\Gamma^2 \left( \frac{1}{3} \right) \, e^{i \, \pi
      /3}}{2^{2/3}},
\end{align}
we obtain
\begin{align}
  \label{GrBj3}
  G_R^{Bj} (k_1, k_2)\big|_{O(1/z_h^4)} \, = \, - \frac{4 \, N_c^2 \,
    \cE_0 \, m^6}{15 \, \Lambda^6} \, \delta^2 ({\un k}_1 + {\un k}_2)
  \, \frac{N}{m^{2/3}_{\bot} \, (1+ \cosh \Delta y)^{1/3}}.
\end{align}
The corresponding two-glueball correlation function scales as
\begin{align}\label{CBj}
  C^{Bj} (k_1, k_2)\big|_{|\Delta y| \gg 1} \sim
  \frac{1}{m^{4/3}_{\bot} \, (\cosh \Delta y)^{2/3}}.
\end{align}
We conclude that rapidity correlations coming from the AdS dual of
Bjorken hydrodynamics are suppressed at large rapidity interval, at
least in the perturbative estimate we have performed. This result
appears to agree with the causality argument
\cite{Gavin:2008ev,Dumitru:2008wn} making appearance of long-range
rapidity correlations unlikely at late times. Moreover, the locality
of $C^{Bj}$ in rapidity suggests that late-time dynamics is not likely
to affect long-range rapidity correlations coming from the early
stages of the collision: hydrodynamic evolution can not ``wash out''
such long-range rapidity correlations.

Note that the complete momentum space two-glueball correlation
function receives contributions from all regions of coordinate space,
i.e., from all $x_1$ and $x_2$. In Sec. \ref{Correlators} we have
calculated the contribution arising from early proper times, while
here we have estimated the late-time contribution. One may expect that
in the complete result the two contributions coming from different
integration regions would simply add together: in such case clearly
the early-time contribution in \eq{Corr_fin} would dominate for large
rapidity intervals, leading to long-range rapidity correlations
arising in the collision.


\section{Summary}
\label{sum}

Let us summarize by first restating that we have found long-range
rapidity correlations in the initial stages of strongly-coupled heavy
ion collisions as described by AdS/CFT correspondence. We expect that
due to causality the correlations would survive the late-time
evolution of the produced medium, though one needs to have a full
solution of the shock wave collision problem to be able to verify this
assertion. The long-range rapidity correlations may be relevant for
the description of the ``ridge'' correlation observed in heavy ion and
proton-proton collisions
\cite{Adams:2005ph,Adare:2008cqb,Alver:2009id,Khachatryan:2010gv}.
Indeed ``ridge'' correlation is characterized not only by the
long-range rapidity correlation, but also by a narrow zero-angle
azimuthal correlation between the triggered and associated particles.
As was suggested in \cite{Gavin:2008ev,Dumitru:2008wn} such azimuthal
correlation may be due to the radial flow of the produced medium. The
advantage of the AdS/CFT approach to the problem is that the full
solution to the problem for a collision of two shock waves with some
non-trivial transverse profiles would have radial flow included in the
evolution of the dual metric, and would be able to demonstrate whether
radial flow is sufficient to lead to the ``ridge'' phenomenon. Indeed
such calculation appears to be prohibitively complicated to do
analytically at the moment.

The correlations we found grow very fast with rapidity interval, as
one can see from \eq{corrf}, while the experimentally observed
correlation
\cite{Adams:2005ph,Adare:2008cqb,Alver:2009id,Khachatryan:2010gv} is
at most flat in rapidity. This result may lead to the conclusion that
the initial stages of heavy ion collisions can not be
strongly-coupled, since this contradicts existing observations. At the
same time, it may happen that higher-order corrections in $\mu_1$ and
$\mu_2$ would affect this rapidity dependence, flattening the
resulting distribution. On yet another hand, such higher-order
corrections become important at later times, and eventually causality
may prohibit further late-time modification to the long-range rapidity
correlations. More work is needed to clarify this important question
about the rapidity-shape of the correlations coming from the solution
of the full problem in AdS.

Assuming that the issue of rapidity shape would be resolved, we would
also like to point out that $k_T$-dependence of obtained correlator
\peq{GR_gen} closely resembles that reported in the data
\cite{Adams:2005ph,Adare:2008cqb,Alver:2009id,Khachatryan:2010gv}: it
starts out growing with $k_T$ at low-$k_T$, and, at higher $k_T$, it
falls off with $k_T$. The location of the maximum of the correlator in
our case was determined by the mass and size of the produced
particles, and was thus energy-independent. It is possible that the
solution of the full problem, resumming all powers of $\mu_1$ and
$\mu_2$ would lead to the maximum of the correlation function given by
$\mu_{1,2}^{1/3}$, which in turn would be inversely proportional to
the thermalization time \cite{Grumiller:2008va,Kovchegov:2009du}, thus
providing an independent way of measuring this quantity. Again more
research is needed to explore this possibility.


\acknowledgments

This research is sponsored in part by the U.S. Department of Energy
under Grant No. DE-SC0004286.



\providecommand{\href}[2]{#2}\begingroup\raggedright\endgroup


\end{document}